\documentclass{IEEEtran}
\usepackage{cite}
\usepackage{amsmath,amssymb,amsfonts}
\usepackage{color}
\usepackage[numbers]{natbib}
\usepackage{multirow}
\usepackage{makecell}
\usepackage[normalem]{ulem}
\PassOptionsToPackage{hyphens}{url}\usepackage{hyperref}
\usepackage{graphicx}
\usepackage[table]{xcolor}
\usepackage{textcomp,nicefrac}
\def\BibTeX{{\rm B\kern-.05em{\sc i\kern-.025em b}\kern-.08em
T\kern-.1667em\lower.7ex\hbox{E}\kern-.125emX}}
\begin{document}
\title{Material Properties of Popular Radiation Detection Scintillator Crystals for Optical Physics Transport Modelling in Geant4}
\author{Lysander Miller, Airlie Chapman, Katie Auchettl, and Jeremy M.\ C.\ Brown
\thanks{This work was supported by The University of Melbourne’s Research Computing Services, the Petascale Campus Initiative, and an Australian Government Research Training Program (RTP) Scholarship.}
\thanks{L.\ Miller and A.\ Chapman are with the Department of Mechanical Engineering, The University of Melbourne, Parkville, VIC 3010, Australia (e-mail: lysanderm@student.unimelb.edu.au; airlie.chapman@unimelb.edu.au).}
\thanks{K.\ Auchettl is with the School of Physics, The University of Melbourne, Parkville, VIC 3010, Australia (e-mail: katie.auchettl@unimelb.edu.au).}
\thanks{J.\ M.\ C.\ Brown is with the Optical Sciences Centre, Department of Physics and Astronomy, Swinburne University of Technology, Hawthorn, VIC 3122, Australia (e-mail: jmbrown@swin.edu.au).}}

\maketitle

\begin{abstract}
Radiation detection is vital for space, medical imaging, homeland security, and environmental monitoring applications. In the past, the Monte Carlo radiation transport toolkit, Geant4, has been employed to enable the effective development of emerging technologies in these fields. Radiation detectors utilising scintillator crystals have benefited from Geant4; however, Geant4 optical physics parameters for scintillator crystal modelling are sparse. This work outlines scintillator properties for GAGG:Ce, CLLBC:Ce, BGO, NaI:Tl, and CsI:Tl. These properties were implemented in a detailed SiPM-based single-volume scintillation detector simulation platform developed in this work. It was validated by its comparison to experimental measurements. For all five scintillation materials, the platform successfully predicted the spectral features for selected gamma ray emitting isotopes with energies between 30 keV to 2 MeV. The full width half maximum (FWHM) and normalised cross-correlation coefficient (NCCC) between simulated and experimental energy spectra were also compared. The majority of simulated FWHM values reproduced the experimental results within a 2\% difference, and the majority of NCCC values demonstrated agreement between the simulated and experimental energy spectra. Discrepancies in these figures of merit were attributed to detector signal processing electronics modelling and geometry approximations within the detector and surrounding experimental environment.
\end{abstract}

\begin{IEEEkeywords}
GAGG, CLLBC, BGO, NaI, CsI, Geant4 optical physics
\end{IEEEkeywords}

\section{Introduction}
\label{sec:introduction}
Radiation detection is important in a wide range of applications such as environmental monitoring, medical imaging, homeland security, and space \citep{Marques2021, Vetter2018}. Scintillator-based radiation detectors offer a low power, lightweight, and accurate means to measure radiation. Geant4 \citep{Agostinelli2003,Allison2006,Allison2016}, a state-of-the-art Monte Carlo radiation transport toolkit that simulates the underlying physical process of interaction and production of particles through matter, has been employed to assist with the design of scintillation detectors \citep{vanderLaan2010, Hartwig2014, Meo2009}. An important aspect of development is accurate scintillator crystal modelling through optimised material properties within Geant4 optical physics. In this work, the material parameters for several common scintillator crystals were used as inputs for a Geant4 (v.\ 11.1.2) simulation platform designed to model scintillator crystals coupled to a silicon photomultiplier (SiPM) based radiation detector. The key material parameters identified to accurately simulate the scintillator crystals were: (1) density, (2) elemental composition, (3) refractive index, (4) optical yield (optical photons produced per MeV of energy deposited), (5) emission spectrum, (6) absorption length, (7) optical decay times, (8) resolution scale, and (9) dimensions \citep{Agostinelli2003}.

The response of five different single-volume scintillator crystal materials under the irradiation of gamma ray emissions from five radioactive isotopes was explored. Scintillator crystals GAGG:Ce, CLLBC:Ce, BGO, NaI:Tl, and CsI:Tl were selected as they offer a range of desirable properties for the applications mentioned previously.  For example, CLLBC:Ce has a high energy resolution \citep{Shirwadkar2012}, BGO provides a high sensitivity \citep{Brunner2017}, GAGG:Ce offers a high sensitivity and good energy resolution \citep{Sibczynski2017}, and NaI:Tl and CsI:Tl are cost effective \citep{Brown2021}. Radioactive sources $^{109}$Cd (1 $\mu$Ci), $^{57}$Co (1 $\mu$Ci), $^{137}$Cs (0.1 $\mu$Ci), $^{22}$Na (1 $\mu$Ci), and $^{152}$Eu (0.5 $\mu$Ci) from Spectrum Techniques were considered in this study as they provided a 30 keV to 2 MeV energy range where major spectral features such as photopeaks, Compton edges, backscatter peaks, and plateaus were investigated. The simulation platform was experimentally benchmarked by comparing the energy spectrum obtained for each scintillator crystal and isotope combination using the full width half maximum (FWHM) and normalised cross correlation coefficient (NCCC) as the figures of merit.

\section{Method}
\subsection{Experimental Platform and Acquisition of Radiation Energy Spectra}\label{sec:experimental_platform}
An off-the-shelf SiPM-3000 from Bridgeport Instruments was chosen as the optical photon detection platform for this work \citep{SiPM3000}. Motivated by the previous applications, the SiPM-3000 was desirable as it was operated using open-source interface software and was composed of a rugged detector housing, read-out electronics, and a Broadcom AFBR-S4N66C013 SiPM array, which had a maximum photodetection efficiency (PDE) of over 55\% \citep{Broadcom2023}. Scintillator crystals GAGG:Ce, CLLBC:Ce, BGO, NaI:Tl, and CsI:Tl were optically bonded to the SiPM array in the SiPM-3000 using 1 mm thick EJ-560 optical pads. Table (\ref{table:scintillator_crystal_types}) summarises the effective atomic number (Z$_{\text{eff}}$), mass attenuation coefficient, dimensions, reflective materials, crystal housing, optical window, and manufacturer for each scintillator crystal.

\begin{table*}
\caption{Scintillator material properties for GAGG:Ce, CLLBC:Ce, BGO, NaI:Tl, and CsI:Tl. The absorption length for CLLBC:Ce was unavailable in the literature. However, Brown \textit{et al}.\ illustrated that it was negligible for a small crystal volume \citep{Brown2023}. Where applicable, the resolution scale was calculated from the energy resolution. The refractive index, emission spectrum, and absorption length data is available in Figs.\ (\ref{fig:material_property_data}a) and (\ref{fig:material_property_data}b) in the appendix.}\label{table:scintillator_crystal_types}
\setlength{\tabcolsep}{6.2pt}
\begin{tabular}
{|c||c|c|c|c|c|c|c|c|c|c|}
\hline
Material & GAGG:Ce & CLLBC:Ce & BGO & NaI:Tl & CsI:Tl \\
\hline
Density (g/cm$^3$) & 6.63 \citep{Kobayashi2012} & 4.06 \citep{Shirwadkar2012, Hawrami2016}  & 7.13 \citep{Mao2008} & 3.67 \citep{Mao2008} & 4.51 \citep{Mao2008}\\
\hline
\makecell[c]{Elemental \\ composition} & \makecell[c]{Gd$_3$Al$_2$Ga$_3$O$_{12}$ \\ (1\% Ce)} & \makecell[c]{Cs$_2$LiLaBr$_{4.8}$Cl$_{1.2}$ \\ (2\% Ce)} & Bi$_4$Ge$_3$O$_{12}$ & NaI (6.5\% Tl) & CsI (0.08\% Tl)\\
\hline
\makecell[c]{Z$_{\text{eff}}$} & 54.4 \citep{Kobayashi2012}  & 47.0 \citep{Hawrami2016} & 74.0 \citep{AdvatechND} & 49.7 \citep{Sibczynski2017}& 54.0 \citep{Sibczynski2017} \\
\hline
\makecell[c]{Mass attenuation \\ coefficient \\ (10$^{-2}$ cm$^2$/g)} & \makecell[c]{7.984 \citep{Berger2010} \\ (at 662 keV)} & \makecell[c]{7.490 \citep{Berger2010} \\ (at 662 keV)} & \makecell[c]{9.979 \citep{Berger2010} \\ (at 662 keV)} & \makecell[c]{7.868 \citep{Berger2010} \\ (at 662 keV)} & \makecell[c]{7.756 \citep{Berger2010} \\ (at 662 keV)}\\
\hline
Refractive index & See Fig.\ (\ref{fig:material_property_data}a) & See Fig.\ (\ref{fig:material_property_data}a) & See Fig.\ (\ref{fig:material_property_data}a) & See Fig.\ (\ref{fig:material_property_data}b) & See Fig.\ (\ref{fig:material_property_data}b)\\
\hline
\makecell[c]{Optical yield \\ (photons/MeV)} & 50,000 \citep{Brown2023a,Brown2021} & 45,000 \citep{Hawrami2016} & 8,500 \citep{Moszynski1997} & 41,000 \citep{Brown2023a,Brown2021} & 61,000 \citep{Moszynski1997}\\
\hline
\makecell[c]{Emission spectrum} & See Fig.\ (\ref{fig:material_property_data}a) & See Fig.\ (\ref{fig:material_property_data}a)& See Fig.\ (\ref{fig:material_property_data}a) & See Fig.\ (\ref{fig:material_property_data}b) & See Fig.\ (\ref{fig:material_property_data}b)\\
\hline
\makecell[c]{Absorption length} & See Fig.\ (\ref{fig:material_property_data}a) & N/A \citep{Brown2023} & See Fig.\ (\ref{fig:material_property_data}a) & See Fig.\ (\ref{fig:material_property_data}b) & See Fig.\ (\ref{fig:material_property_data}b)\\
\hline
\makecell[c]{Optical decay \\ time(s) (ns)} & \makecell[c]{87 (90\%), \\ 255 (10\%) \citep{Brown2023a,Brown2021}} & \makecell[c]{130 (82.5\%), \\ 784 (17.5\%) \citep{Brown2023}} & \makecell[c]{317 (100\%) \citep{EPICBGO2023}} & \makecell[c]{220 (96\%), \\ 1500 (4\%) \citep{Brown2023a,Brown2021}} & \makecell[c]{1000 (100\%) \citep{Brown2023a,Brown2021}}\\
\hline
Resolution scale & \makecell[c]{3.08 \citep{Brown2023a,Brown2021} \\ (at 511 keV)} & \makecell[c]{2.13 \citep{Hawrami2016} \\ (at 662 keV)} & \makecell[c]{3.80 \citep{EPICBGO2023} \\ (at 662 keV)} & \makecell[c]{3.50 \citep{Brown2023a,Brown2021} \\ (at 511 keV)} & \makecell[c]{4.87 \citep{GrodzickaKobylka2017} \\ (at 662 keV)}\\
\hline
Crystal dimensions & 25.4 mm cube& \makecell[c]{24.98 mm diameter \\ 25.33 mm height} & 25.4 mm cube & 25.4 mm cube& 25.4 mm cube\\
\hline
\makecell[c]{Reflective material} & \makecell[c]{0.015 mm EPO-TEK-301 \\ 0.065 mm ESR} & \makecell[c]{1 mm Teflon sides \\ 1.5 mm GORE top} & \makecell[c]{0.015 mm EPO-TEK-301 \\ 0.065 mm ESR} & 1 mm Teflon & 1 mm Teflon\\
\hline
\makecell[c]{Crystal housing} & -- & \makecell[c]{2.5 mm neoprene \\ 1 mm 2A12 Al} & -- & \makecell[c]{1 mm 2A12 Al} & \makecell[c]{1 mm 2A12 Al}\\
\hline
\makecell[c]{Optical window} & -- & \makecell[c]{1 mm EJ-560 \\ 0.25 mm glass} & -- & \makecell[c]{2 mm EPO-TEK-301 \\ 2 mm glass} & \makecell[c]{2 mm EPO-TEK-301 \\ 2 mm glass}\\
\hline
Manufacturer & Epic Crystal & RMD & Epic Crystal & Epic Crystal & Epic Crystal\\
\hline
\end{tabular}
\end{table*}

\begin{table}
\caption{Optical material properties for the non-scintillator components in the Geant4 simulation platform. Refractive index, reflectivity, and absorption length data is available in Figs.\ (\ref{fig:material_property_data}c) and (\ref{fig:material_property_data}d) in the appendix.}\label{fig:non_scintillator_material_properties}
\setlength{\tabcolsep}{4.3pt}
\begin{tabular}{|c||c|c|c|c|}
\hline
Material & \makecell[c]{Density \\ (g/cm$^3$)} & \makecell[c]{Elemental \\ composition} & \makecell[c]{Refractive index, \\ reflectivity, and \\ absorption length} & Reference\\
\hline
EPO-TEK-301 & 1.2 & HCO & See Fig.\ (\ref{fig:material_property_data}c) & \citep{EPOTEK3012013} \\
\hline
Teflon tape & 2.2& C$_2$F$_4$& See Fig.\ (\ref{fig:material_property_data}d) & \citep{Janecek2012}\\
\hline
ESR & 1.29 & H$_8$C$_{10}$O$_4$& See Fig.\ (\ref{fig:material_property_data}d) & \citep{3M2020}\\
\hline
\makecell[c]{GORE diffuse \\ reflector} & 0.65 & C$_2$F$_4$& See Fig.\ (\ref{fig:material_property_data}d) & \citep{Janecek2012} \\
\hline
Glass & 2.203 & SiO$_2$ & See Fig.\ (\ref{fig:material_property_data}c) & \citep{Brown2020}\\
\hline
SiPM pixel & 2.33 & Si & See Fig.\ (\ref{fig:material_property_data}c) & \citep{Philipp1960}\\
\hline
\makecell[c]{EJ-560 \\ optical pad} & 1.03& H$_6$C$_2$OSi & See Fig.\ (\ref{fig:material_property_data}c) & \citep{EJ5602021}\\
\hline
\end{tabular}
\end{table}

\begin{figure}[h!]
\includegraphics[width=0.46\linewidth]{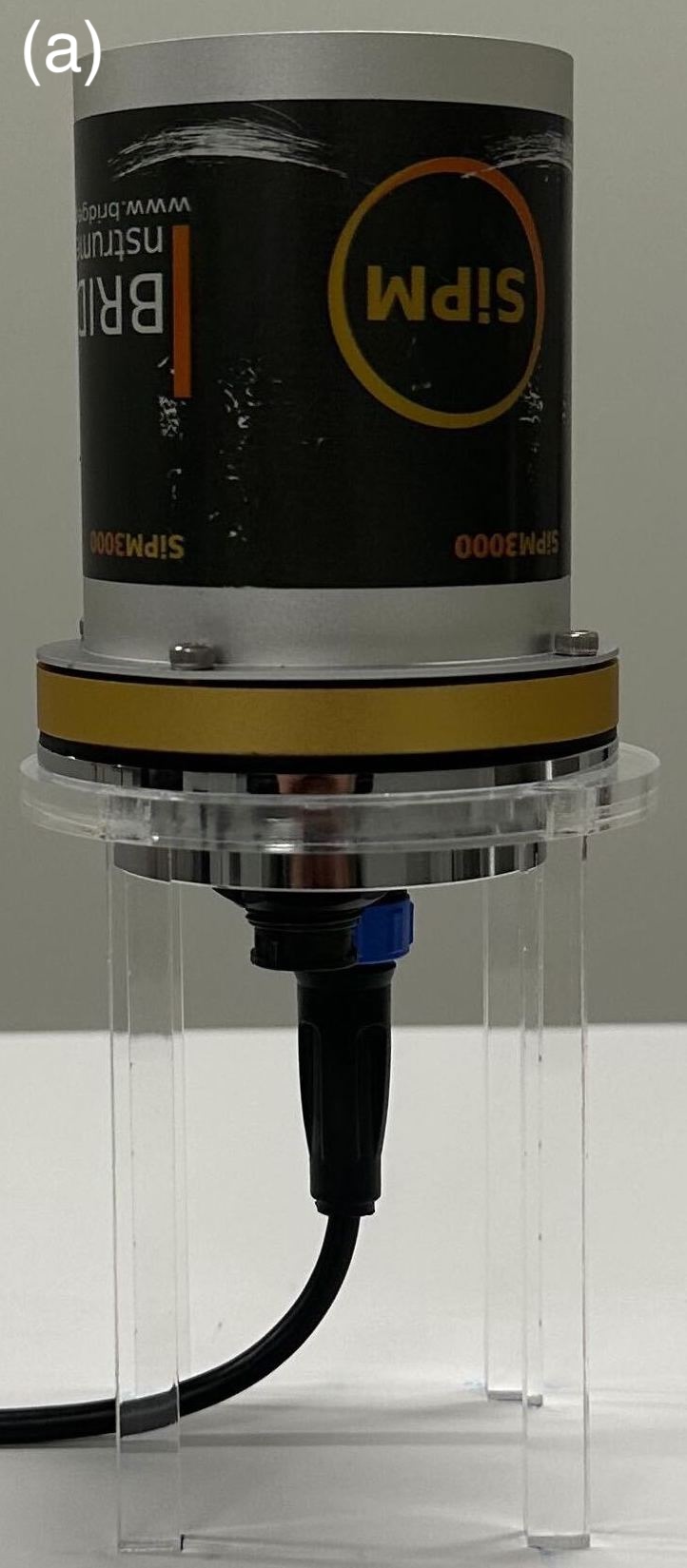}
\includegraphics[width=0.512\linewidth]{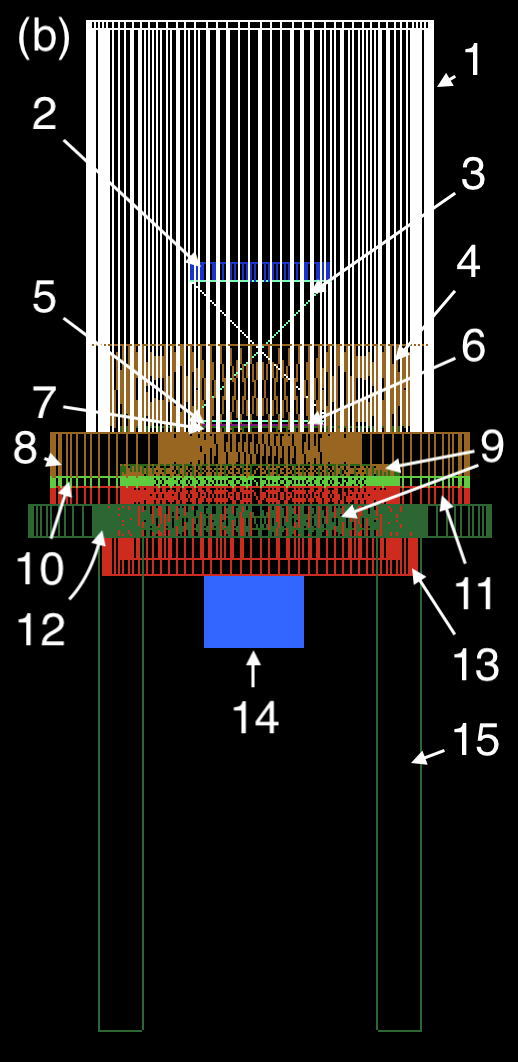}
\caption{Experimental (a) and simulated (b) SiPM-based scintillation detector platforms. With reference to the simulation platform geometries outlined in Table (\ref{table:geant4_geometries}), the 15 components are: (1) detector cover, (2) radioactive check source, (3) scintillator, (4) lip, (5) optical pad, (6) SiPM array, (7) SiPM array PCB, (8) fixture, (9) internal PCBs, (10) gasket, (11) join, (12) detector holder base, (13) bulk, (14) plug and pin connectors, and (15) detector holder legs.}
\label{fig:complete_housing}
\end{figure}

The system was placed in a custom-made detector holder (see Fig.\ (\ref{fig:complete_housing}a)), which was centered on a table away from the lab walls to minimise the chance of Compton scattering with the surrounding environment. One of each radioactive source ($^{109}$Cd, $^{57}$Co, $^{137}$Cs, $^{22}$Na, and $^{152}$Eu) was placed on top of the scintillator crystal within the detector housing to further minimise scattering. After allowing an hour for the SiPM-3000 to thermalise, the detector was calibrated for each crystal type. The integration time, dead time, electronic gain, pulse trigger, and noise trigger were optimised to minimise the noise and low-level detection threshold. For this work, the low-level detection threshold was defined as the minimum deposited gamma ray energy where the corresponding optical photon count exceeded 5\% of the primary photopeak count to minimise the impact of detector noise on the threshold. An integration time of 2 $\mu$s was set for GAGG:Ce, BGO, and NaI:Tl, whereas CLLBC:Ce and CsI:Tl used 3 $\mu$s due to their longer optical decay times (see Table (\ref{table:scintillator_crystal_types})). The SiPM-3000 operating voltage was fixed at 33 V (default), yielding an overvoltage of 4.5 V for the SiPM array. For each scintillator crystal and radioactive source combination, data acquisition ran over 30 minutes in real time and at room temperature (20$^\circ$C) to reduce statistical fluctuations in the energy spectrum and ensure each photopeak could be resolved. Background measurements were also recorded at room temperature over 30 minutes in real time then subtracted from each energy spectrum during post-processing.

\subsection{Geant4 Simulation Platform Geometry, Materials, and Optical Data Tables}
The Geant4 simulation platform was designed to emulate the experimental platform described in Sec.\ (\ref{sec:experimental_platform}). Scintillator crystal dimensions, material properties, and encapsulation implemented in the simulation platform are summarised in Table (\ref{table:scintillator_crystal_types}) with reference to scintillator refractive index, emission spectrum, and absorption length data in Figs.\ (\ref{fig:material_property_data}a) and (\ref{fig:material_property_data}b) in the appendix. The SiPM array and optical pad dimensions were consistent with the experimental platform. The optical material properties for the SiPMs, optical pad, and reflective materials implemented in the simulation platform are displayed in Table (\ref{fig:non_scintillator_material_properties}) with refractive index, reflectivity, and absorption length presented in Figs.\ (\ref{fig:material_property_data}c) and (\ref{fig:material_property_data}d) in the appendix. SiPM-3000 read-out electronics, detector housing, detector holder, radioactive check source, table, and lab environment (walls and floor) were included in the Geant4 application. The description, dimensions, and materials for each geometry are summarised in Table (\ref{table:geant4_geometries}). In the simulation platform, the centre of the first and second lab walls were positioned at ($x=$1710 mm, $y=-$1205 mm) and ($-$1980 mm, $-$430 mm) relative to the centre of the table. Other lab components, e.g., the remaining walls, were disregarded as their contribution of Compton scattering was assumed to be negligible as they were further away from the detector. Surrounding environmental air was modelled using the built-in Geant4 material G4$\_$AIR \citep{Geant4Collab2022}. An OpenGL wire trace visualisation of the detector, detector holder, and radioactive check source implemented in Geant4 is shown in Fig.\ (\ref{fig:complete_housing}b).

\begin{table}[h!]
\caption{Component, dimensions, and material properties of the SiPM-3000 and surrounding geometries implemented in the Geant4 application. Material density is represented by $\rho$. The epoxy was composed of HCO with a density of 1.2 g$/$cm$^3$.}\label{table:geant4_geometries}
\setlength{\tabcolsep}{3.5pt}
\begin{tabular}{|c|c||c|c|}
\hline
\makecell[c]{Geometry} & \makecell[c]{Component} & \makecell[c]{Dimensions} & \makecell[c]{Materials} \\
\hline
\multirow{37}{*}{SiPM-3000} & \multirow{4}{*}{SiPM}&\makecell[c]{Box (x, y, z): \\ 6.14, 6.14, 0.1 (mm)}&\makecell[c]{Si \\ $\rho = 2.33$ g$/$cm$^3$}\\
\cline{3-4}
& &\makecell[c]{Box (x, y, z): \\ 6.14, 6.14, 0.2 (mm)}&\makecell[c]{Glass \\ SiO$_2$ \\ $\rho = 2.203$ g$/$cm$^3$}\\
\cline{2-4}
& \makecell[c]{SiPM \\ array}&\makecell[c]{Box (x, y, z): \\ 25.52, 25.52, 0.3 (mm)}&\makecell{4 x 4 array \\ of SiPMs}\\
\cline{2-4}
& \makecell[c]{SiPM \\ array \\ PCB}&\makecell[c]{Disk (radius, z): \\ 25.4, 1 (mm)}&\multirow{1}{*}{\makecell[c]{FR4 \\ Glass:Epoxy \\ 52.8\%:47.2\% \\ $\rho = 1.86$ g$/$cm$^3$}}\\
\cline{2-3}
& \makecell[c]{Internal \\ PCBs \\ (2x)} &\makecell[c]{Disk (radius, z): \\ 25.4, 2 (mm)}&\\
\cline{2-4}
& Gasket &\makecell[c]{Annulus \\ (thickness, radius, z): \\ 12.5, 38, 2 (mm)}&\makecell[c]{Rubber \\ H$_8$C$_4$ \\ $\rho = 0.95$ g$/$cm$^3$}\\
\cline{2-4}
& \makecell[c]{Pin \\ connector} &\makecell[c]{Annulus \\ (thickness, radius, z): \\ 1, 6.5, 12 (mm)}&\multirow{4}{*}{\makecell[c]{Plastic \\ H$_6$C$_6$O$_2$ \\ $\rho = 1.3$ g$/$cm$^3$}}\\
\cline{2-3}
& \makecell[c]{Plug \\ connector} &\makecell[c]{Annulus \\ (thickness, radius, z): \\ 3.5, 9, 13 (mm)}&\\
\cline{2-4}
& Fixture &\makecell[c]{Annulus \\ (thickness, radius, z): \\ 20, 38, 8 (mm)}&\multirow{13}{*}{\makecell[c]{Al \\ $\rho = 2.7$ g$/$cm$^3$}}\\
\cline{2-3}
& Lip &\makecell[c]{Annulus \\ (thickness, radius, z): \\ 1, 30.5, 16 (mm)}&\\
\cline{2-3}
& Join &\makecell[c]{Annulus \\ (thickness, radius, z): \\ 12.5, 38, 3 (mm)}&\\
\cline{2-3}
& Bulk &\makecell[c]{Annulus \\ (thickness, radius, z): \\ 3, 28.5, 13 (mm)}&\\
\cline{2-3}
& \makecell[c]{Detector \\ cover} &\makecell[c]{Annulus \\ (thickness, radius, z): \\ 1.25, 31.75, 75 (mm)}&\\
\cline{1-3}
\multirow{5}{*}{Table} & \makecell[c]{Legs \\ (4x)} &\makecell[c]{Box (x, y, z): \\ 26, 26, 91 (mm)}&\\
\cline{2-4}
& Body &\makecell[c]{Annulus \\ (thickness, x, y, z): \\ 18, 520, 520, 708 (mm)}&\makecell[c]{MDF \\ H:C:O \\ 6\%:50\%:44\% \\ $\rho = 0.7$ g$/$cm$^3$}\\
\hline
\multirow{3}{*}{\makecell[c]{Detector \\ holder}} & Base &\makecell[c]{Annulus \\ (thickness, radius, z): \\ 12, 42, 6 (mm)}&\multirow{5}{*}{\makecell[c]{Perspex \\ H$_8$C$_5$O \\ $\rho = 1.18$ g$/$cm$^3$}}\\
\cline{2-3}
& \makecell[c]{Legs \\ (4x)} &\makecell[c]{Box (x, y, z): \\ 6, 8, 90 (mm)}&\\
\cline{1-3}
\makecell[c]{Radioactive \\ check source} & -- &\makecell[c]{Disk (radius, z): \\ 12.7, 3.2 (mm)}&\\
\hline
\multirow{5}{*}{Environment} & Wall 1&\makecell[c]{Box (x, y, z): \\ 1, 1.45, 3 (m)}&\multirow{5}{*}{G4$\_$CONCRETE}\\
\cline{2-3}
& Wall 2&\makecell[c]{Box (x, y, z): \\ 1, 3, 3 (m)}&\\
\cline{2-3}
& Floor&\makecell[c]{Box (x, y, z): \\ world x, world y, 1 m}&\\
\hline
\end{tabular}
\end{table}

\begin{figure}[h!]
\includegraphics[width=0.495\linewidth]{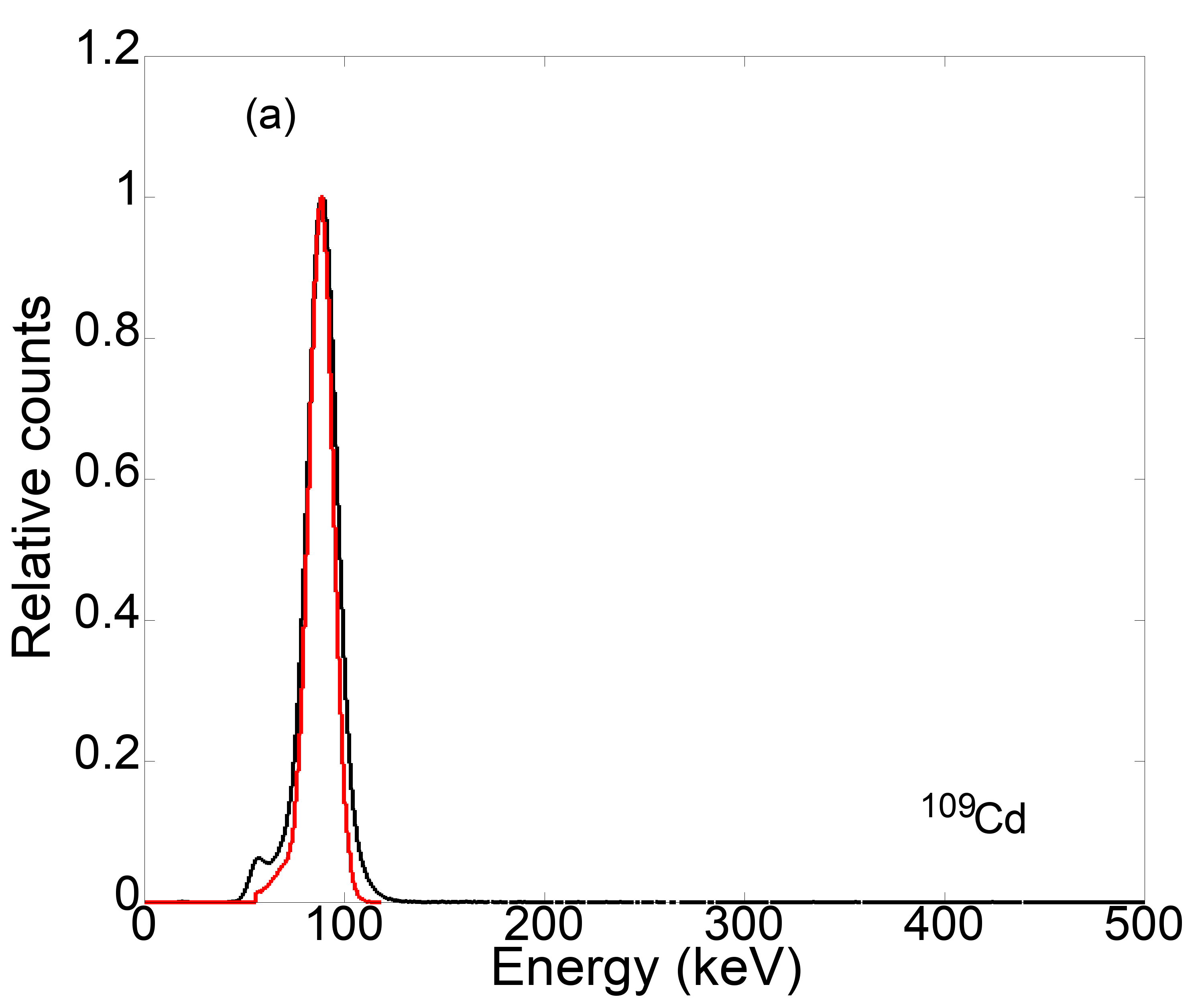}
\includegraphics[width=0.495\linewidth]{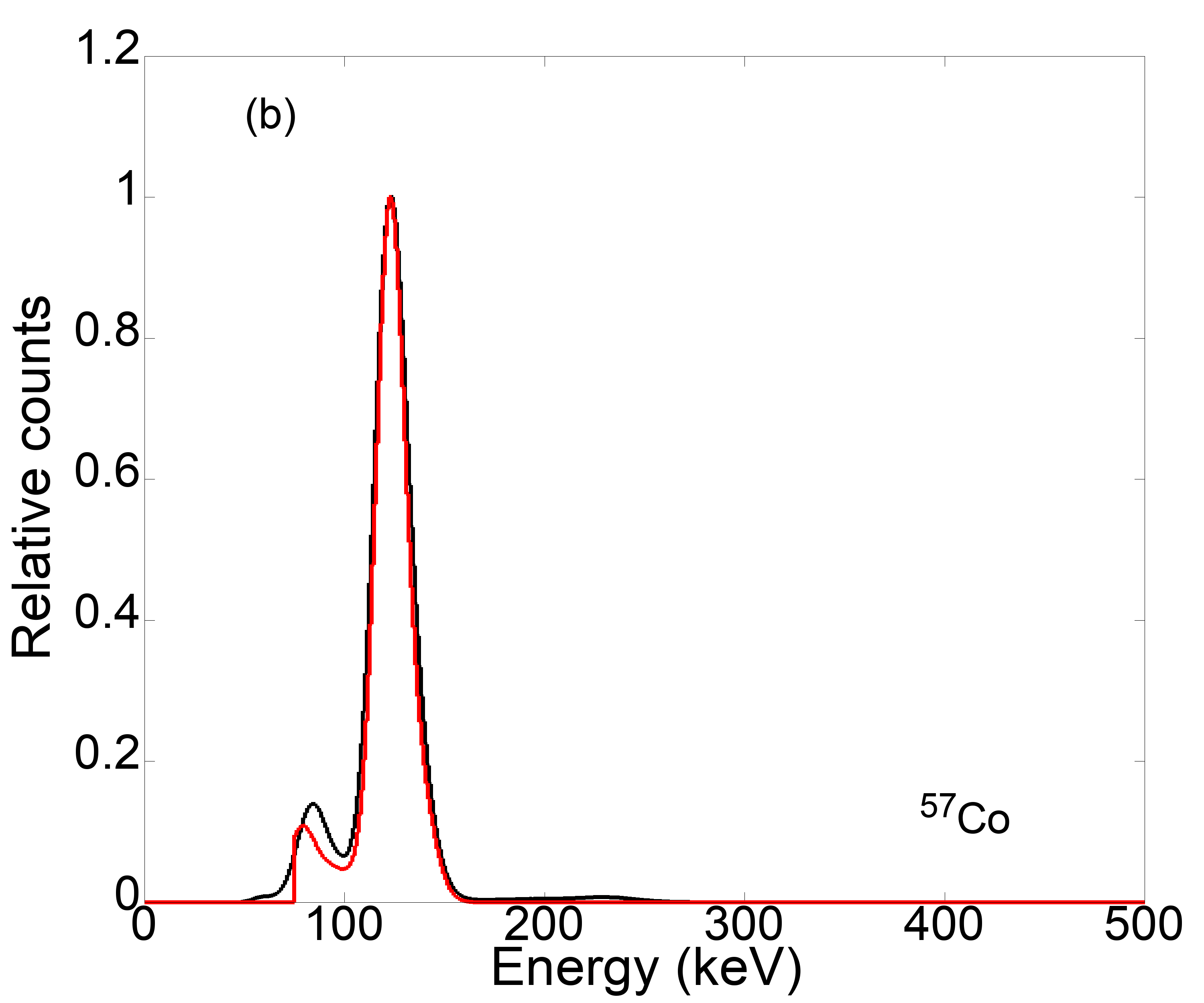}
\centerline{\includegraphics[width=0.495\linewidth]{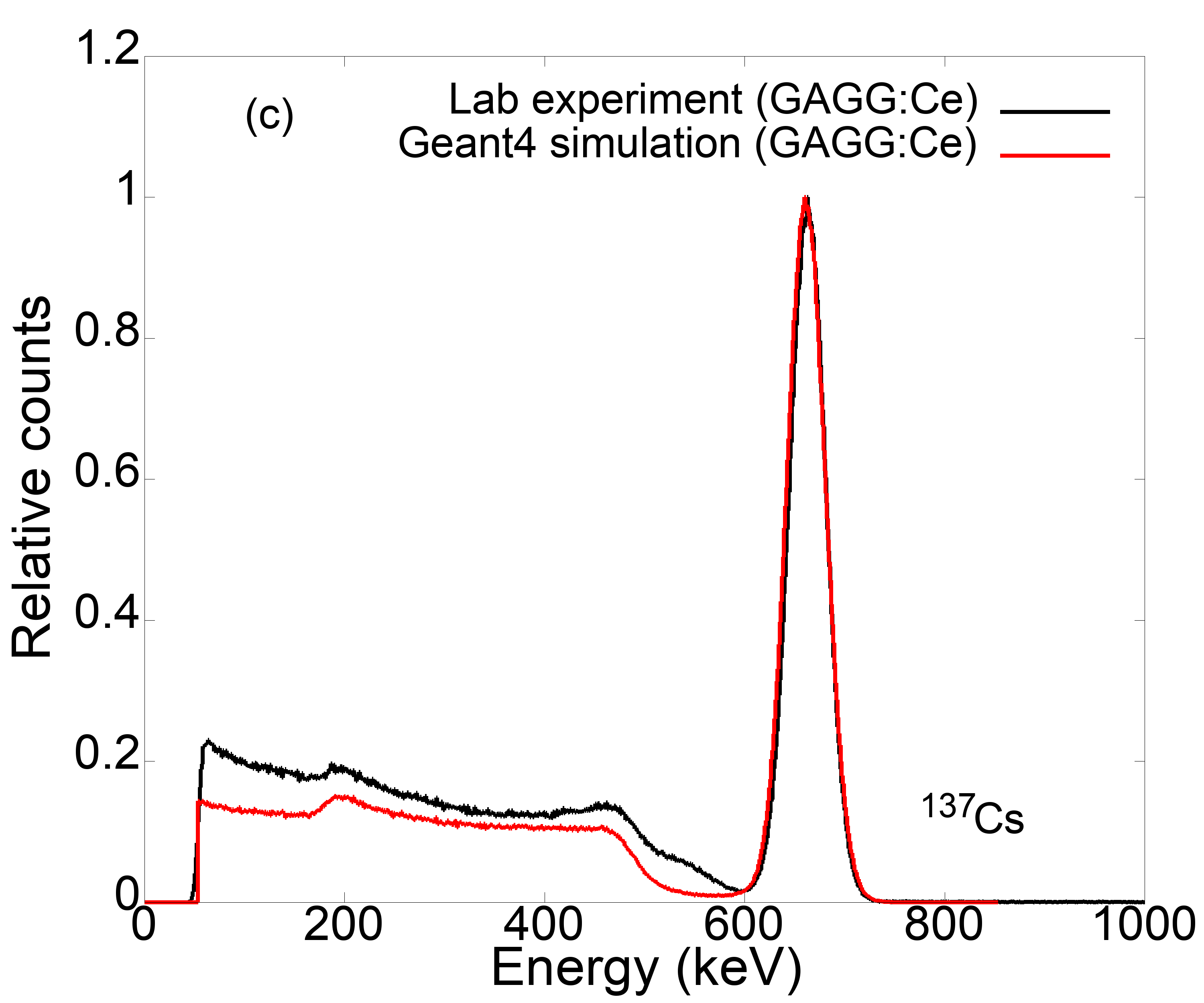}}
\includegraphics[width=0.495\linewidth]{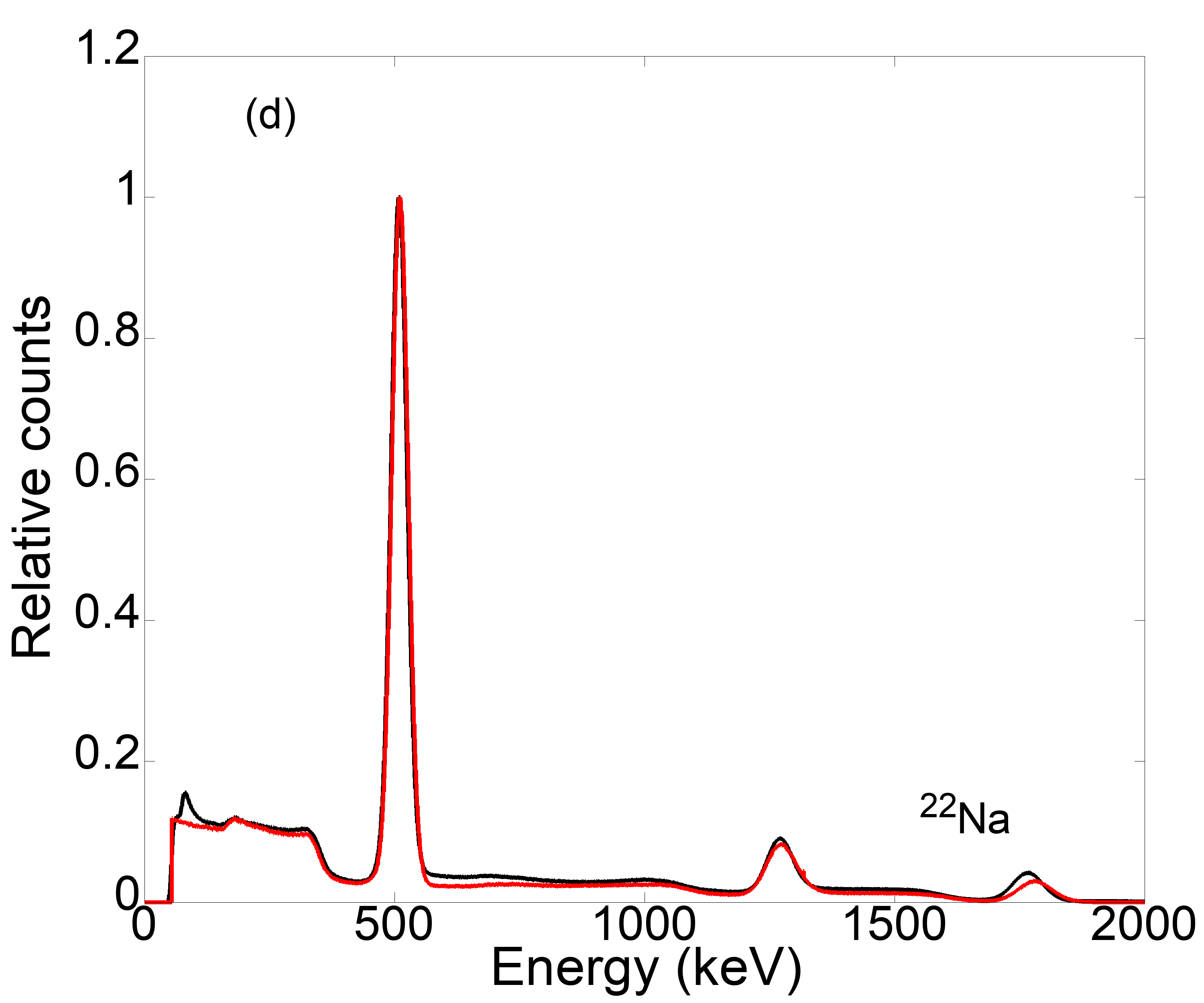}
\includegraphics[width=0.495\linewidth]{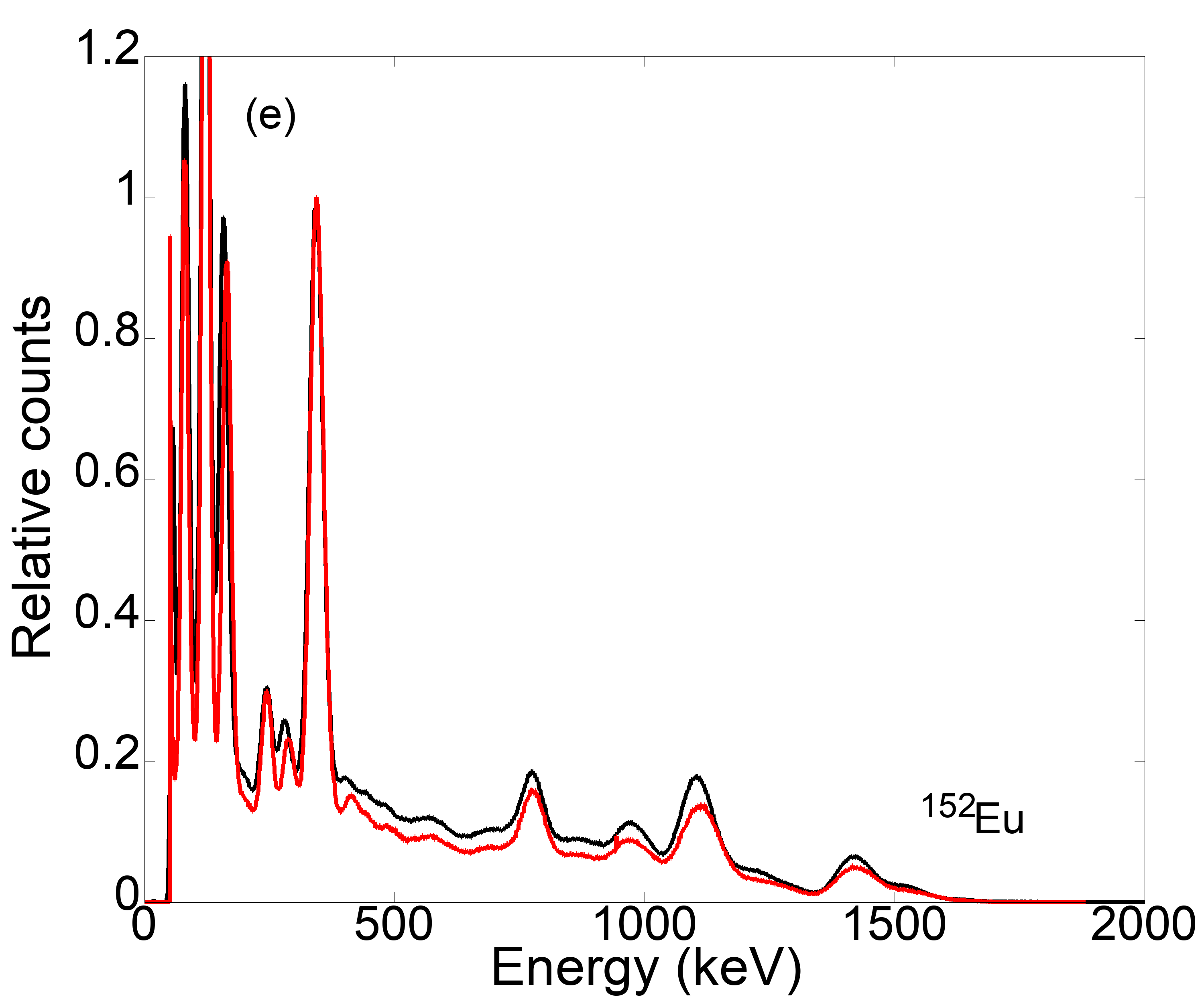}
\caption{Experimental (black line) and simulated (red line) $^{109}$Cd (a), $^{57}$Co (b), $^{137}$Cs (c), $^{22}$Na (d), and $^{152}$Eu (e) gamma ray energy spectra for the GAGG:Ce scintillation detector. The color version is available online.}
\label{fig:gagg}
\end{figure}

\begin{figure}[h!]
\includegraphics[width=0.495\linewidth]{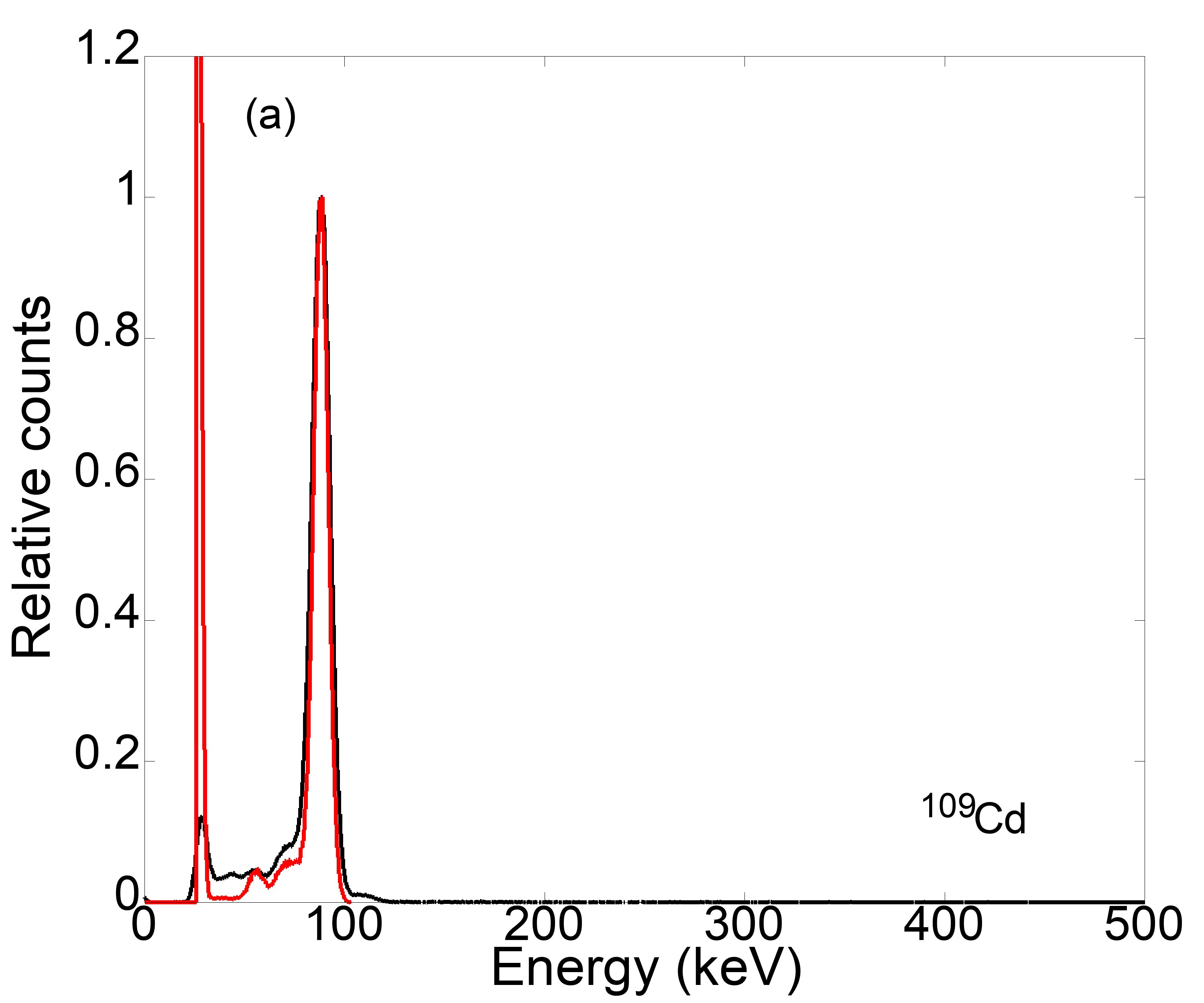}
\includegraphics[width=0.495\linewidth]{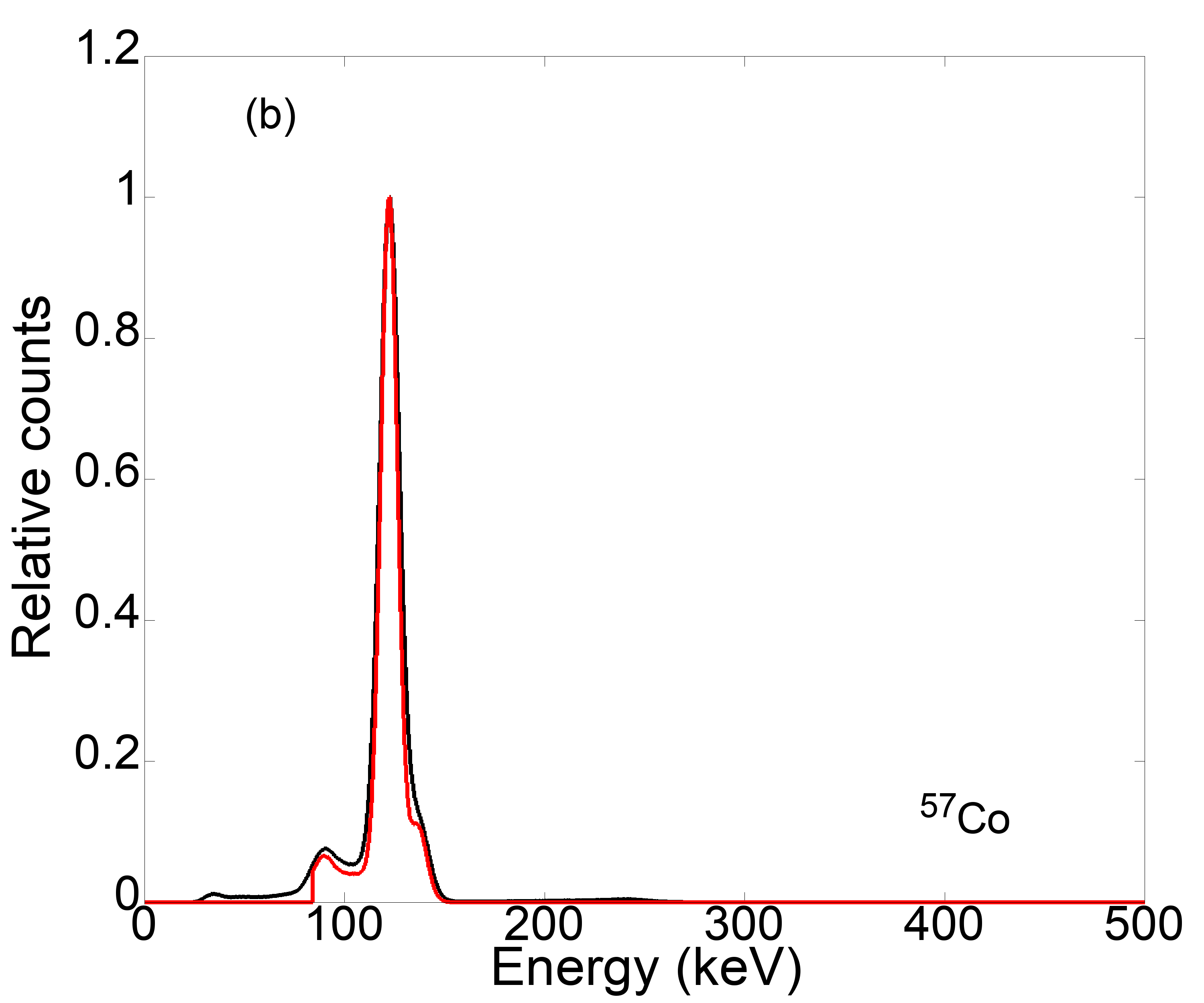}
\centerline{\includegraphics[width=0.495\linewidth]{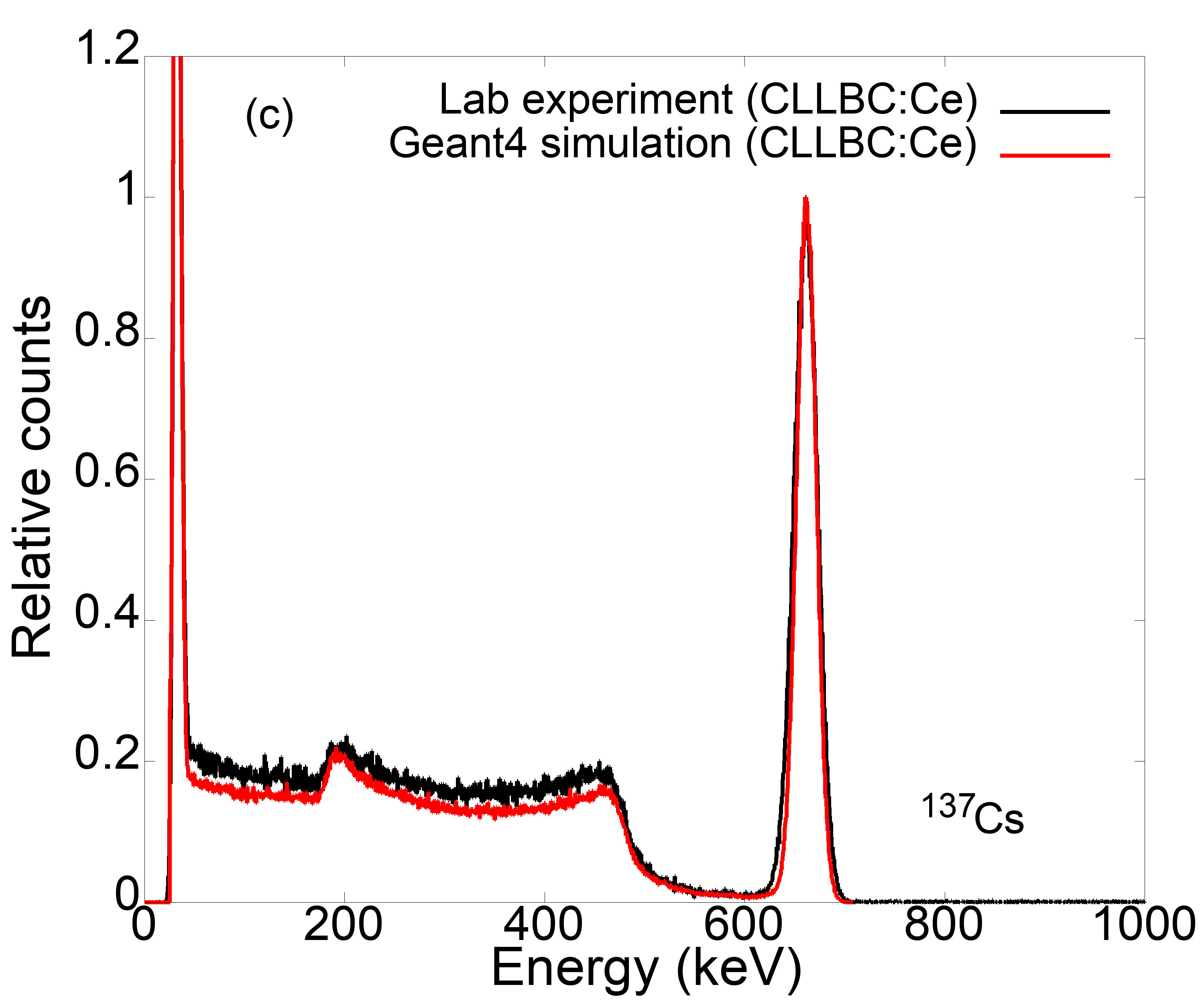}}
\includegraphics[width=0.495\linewidth]{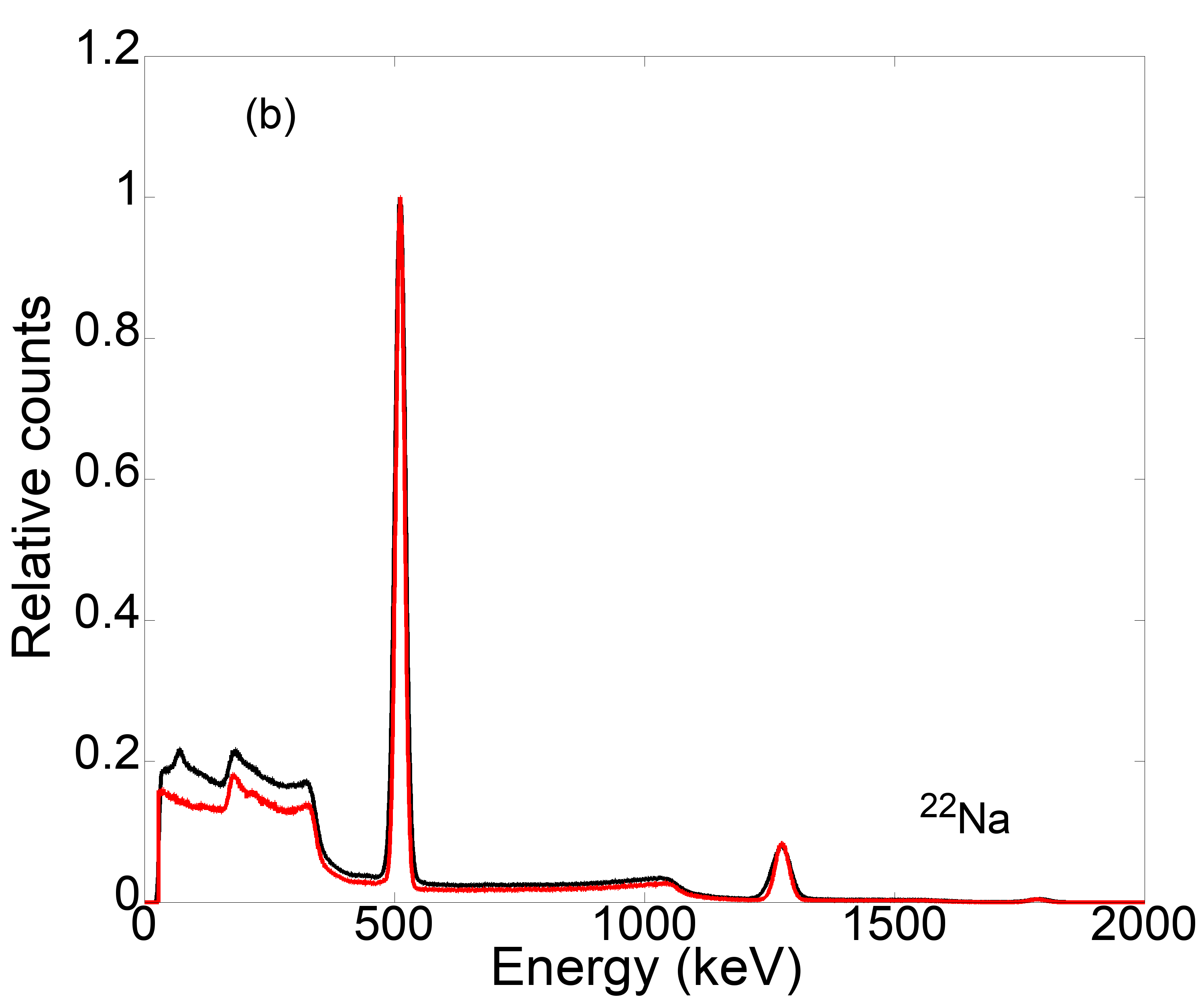}
\includegraphics[width=0.495\linewidth]{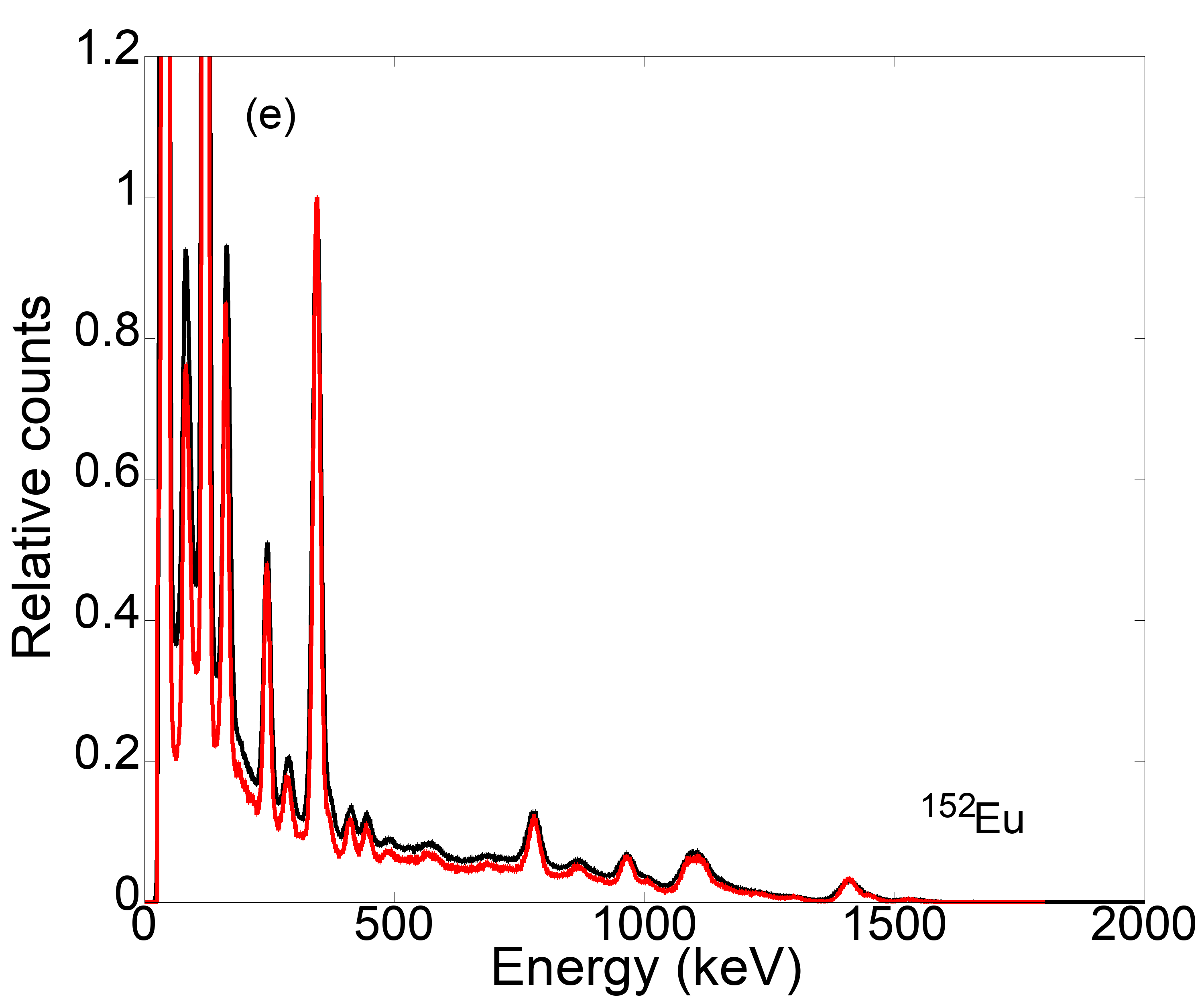}
\caption{Experimental (black line) and simulated (red line) $^{109}$Cd (a), $^{57}$Co (b), $^{137}$Cs (c), $^{22}$Na (d), and $^{152}$Eu (e) gamma ray energy spectra for the CLLBC:Ce scintillation detector. The color version is available online.}
\label{fig:cllbc}
\end{figure}

\subsection{Geant4 Application Physics and Optical Surface Modelling}
The Geant4 Option4 EM physics list (G4EmStandardPhysics\_option4) was used to model the gamma ray, X-ray, and electron transport in the simulation platform \citep{Allison2016, Geant4Collab2022}. The settings for radioactive decay, atomic de-excitation, PIXE, auger electrons, and fluorescence were enabled. Moreover, the particle production length cut was set to 100 $\mu$m as it was the smallest dimension in the simulation platform. Finally, the low energy cut off was fixed at 100 eV to match the low energy limit of the Geant4 physics model \citep{G4PhysRefManual}. Optical photon absorption, reflection, and refraction at optical boundaries were modelled using the Geant4 implementation of the Unified model \citep{Geant4Collab2022, Levin1996}. Optical surfaces were defined in three ways: (1) `dielectric-to-dielectric' with a ground-back-painted finish for surfaces between Teflon and GORE to CLLBC:Ce, NaI:Tl, and CsI:Tl to account for total internal reflection from the air gap, (2) ‘dielectric-to-metal’ with a ground finish for surfaces between ESR and the EPO-TEK-301 glue that binds it to the GAGG:Ce and BGO crystals, and (3) ‘dielectric-to-dielectric’ with a ground finish for the remaining non-reflective material surfaces. All optical surfaces had a 0.1-degree surface roughness as it is not possible to have a completely polished surface \citep{Brown2023, Nilsson2015, vanderLaan2010}.

\subsection{Geant4 Application Validation Simulations and Figures of Merit}
The simulation platform was used to model twenty million radioactive decays for each radioactive source and scintillator crystal combination, totaling 25 simulations. For these simulations, the isotope particle gun was positioned at the centre of each radioactive check source to approximate the isotope location. The optical photons produced by the scintillator crystals were scored by the SiPM array. Optical photon counts were calculated from the scored optical photon data taking into account the integration times and wavelength-dependent PDE post-processing. The integration times were consistent with the experiments as described in Sec.\ (\ref{sec:experimental_platform}), and the PDE data was linearly interpolated for an overvoltage of 4.5 V using the 3.5 V and 8 V overvoltage from the AFBR-S4N66C013 SiPM array data sheet \citep{Broadcom2023}. Energy calibration was implemented by least squares fitting photopeak centroids in the optical photon count domain to the corresponding gamma ray energy and setting the minimum energy for the simulated data as the experimental low-level detection threshold. Quadratic functions were fit to the centroids of $^{109}$Cd, $^{137}$Cs, and $^{22}$Na for GAGG:Ce, CLLBC:Ce, NaI:Tl, and CsI:Tl. A linear function was fit to the $^{137}$Cs and $^{22}$Na centroids for BGO as the $^{109}$Cd centroid was filtered by the low-level detection threshold.

Two figures of merit were selected to benchmark the Geant4 application: the FWHM and NCCC. The FWHM was used to quantify the performance of photopeaks in the simulation platform. It was calculated from the standard deviation of a Gaussian function with a quadratic background that was fit to each photopeak using linear least squares. The NCCC was used to quantify the overall performance of the simulated energy spectrum $E_{\text{sim}}(i)$ compared to the experimental energy spectrum $E_{\text{exp}}(i)$ over all energy bins $i$:
\begin{equation}
\text{NCCC} = \frac{|\sum_{i=0}^nE_{\text{sim}}(i)E_{\text{exp}}(i)|}{|\sum_{i=0}^nE_{\text{sim}}^2(i)|^{1/2}|\sum_{i=0}^nE_{\text{exp}}^2(i)|^{1/2}},\label{eq:nccc}
\end{equation}
where values greater than 0.99 indicate an acceptable fit \citep{HernandezAndres1998}. For the NCCC calculations, the experimental energies were used as the $n+1$ spectral channels in (\ref{eq:nccc}) and the corresponding simulation counts were determined via interpolation.

\begin{figure}[h!]
\includegraphics[width=0.495\linewidth]{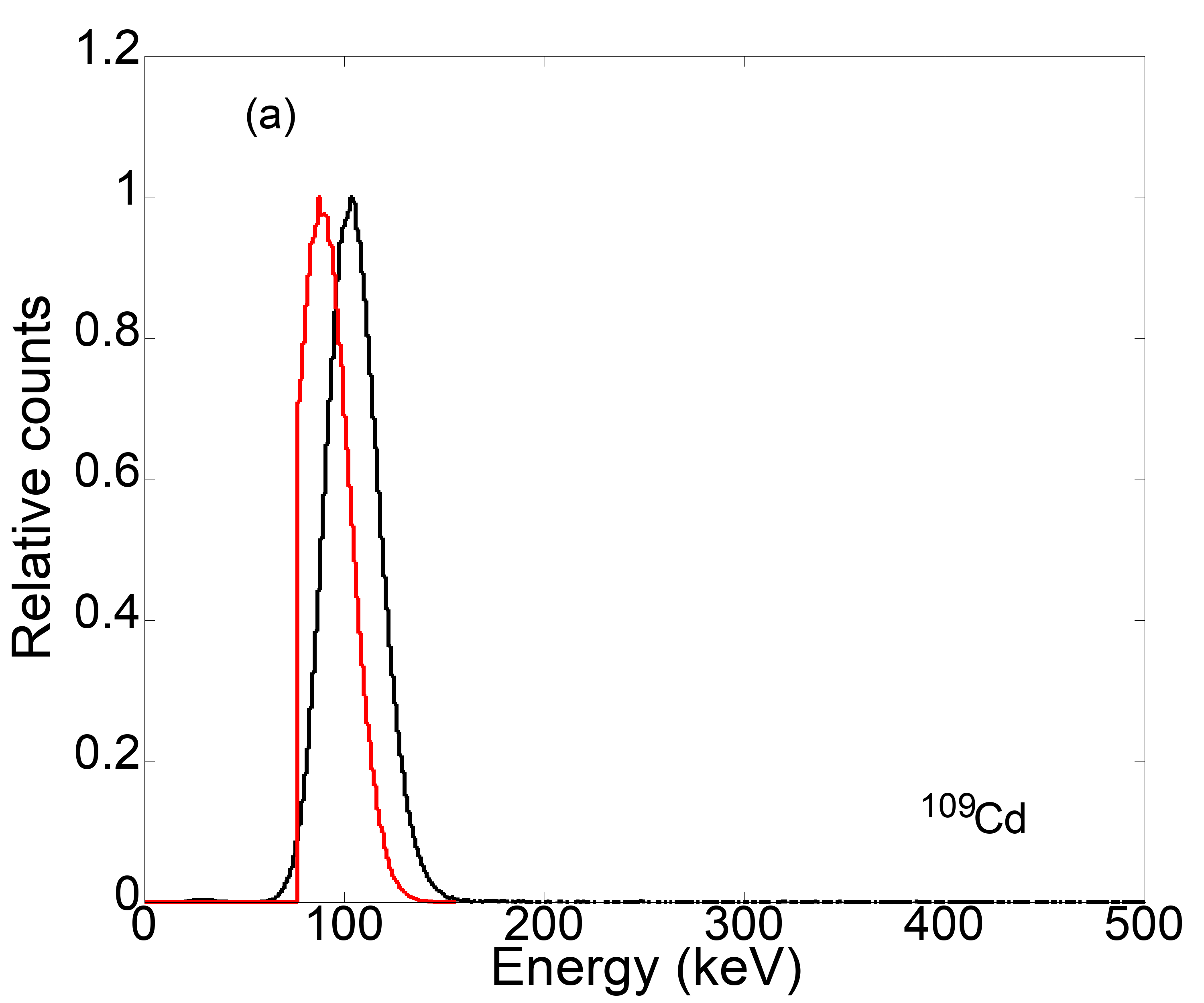}
\includegraphics[width=0.495\linewidth]{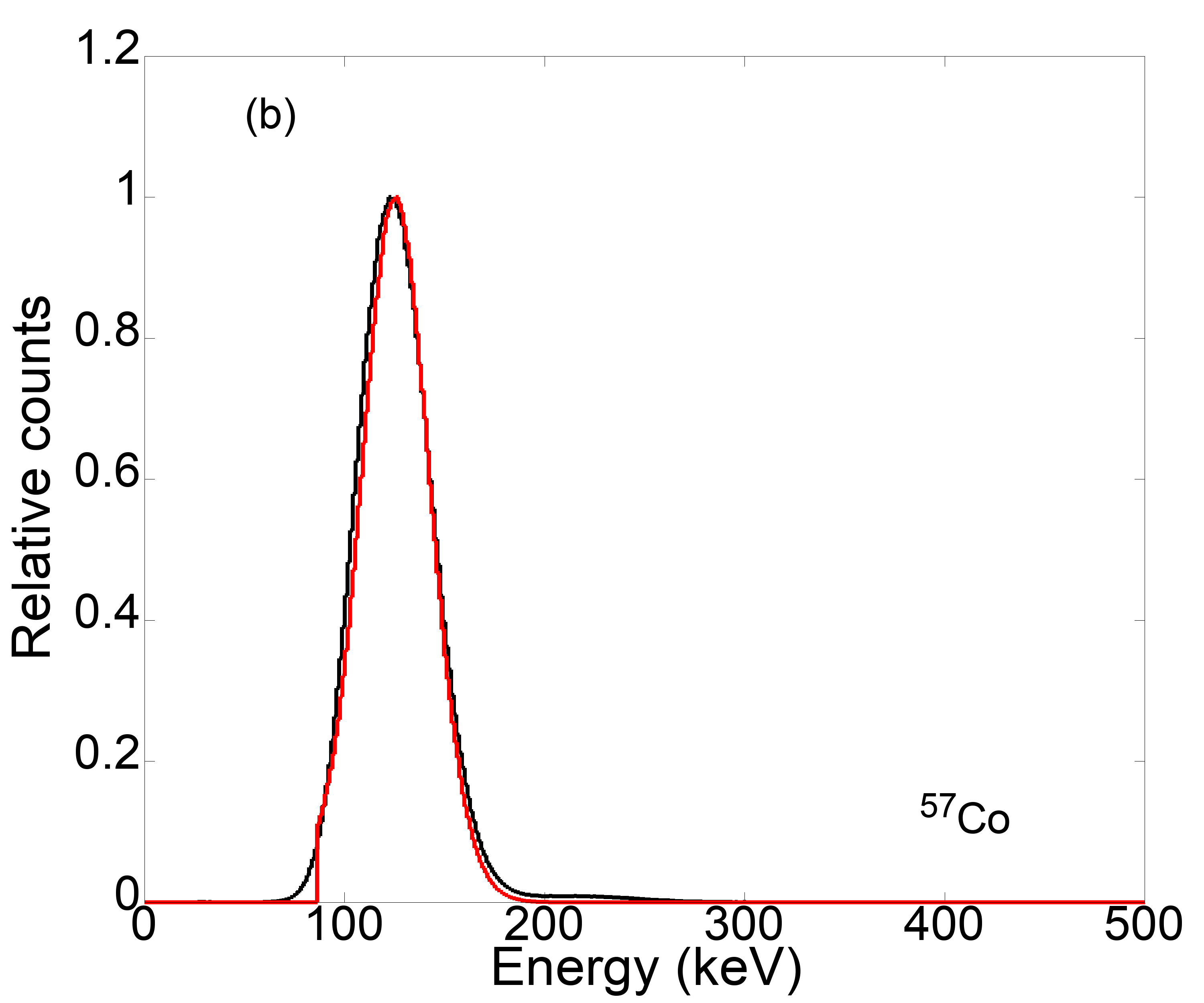}
\centerline{\includegraphics[width=0.495\linewidth]{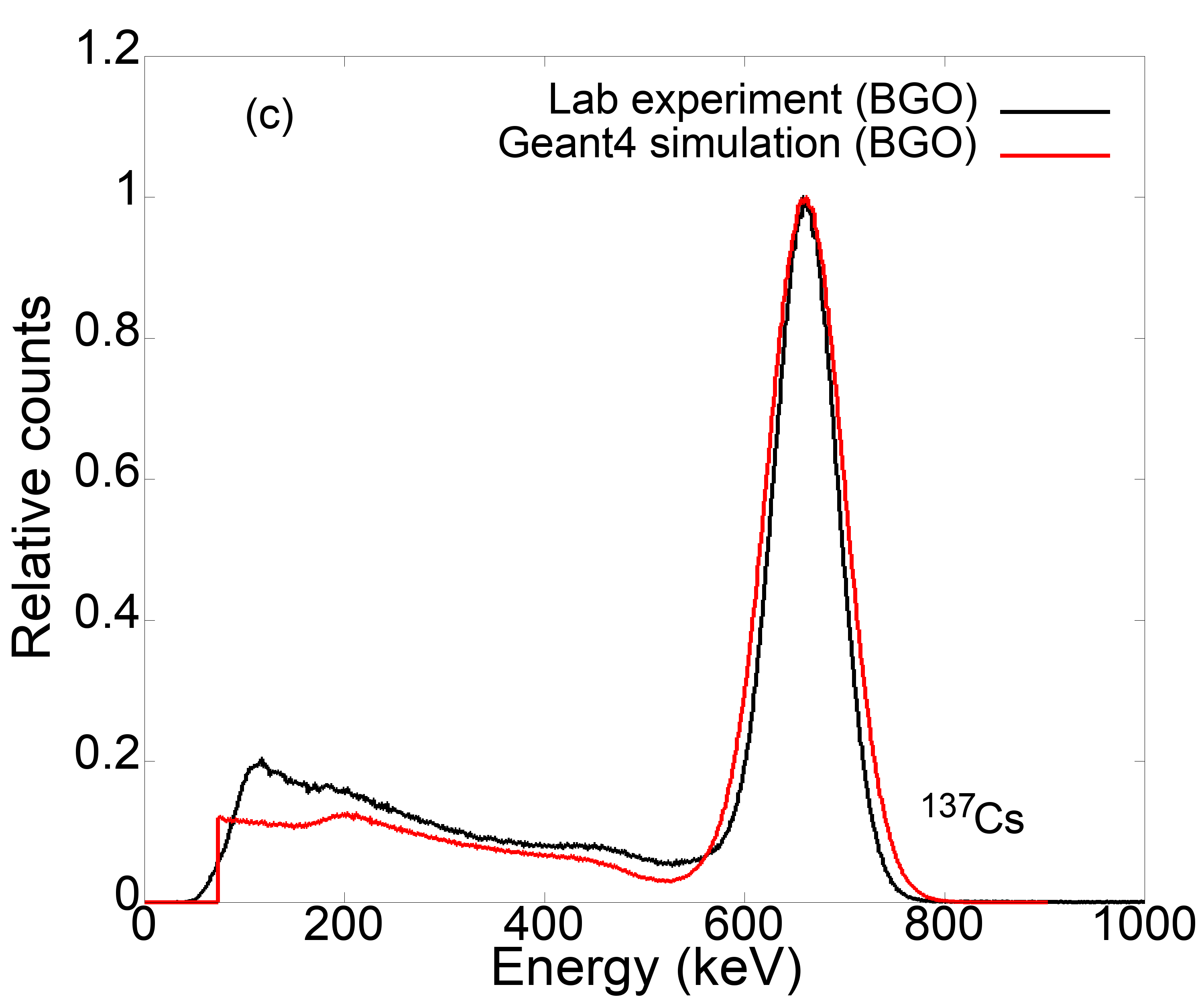}}
\includegraphics[width=0.495\linewidth]{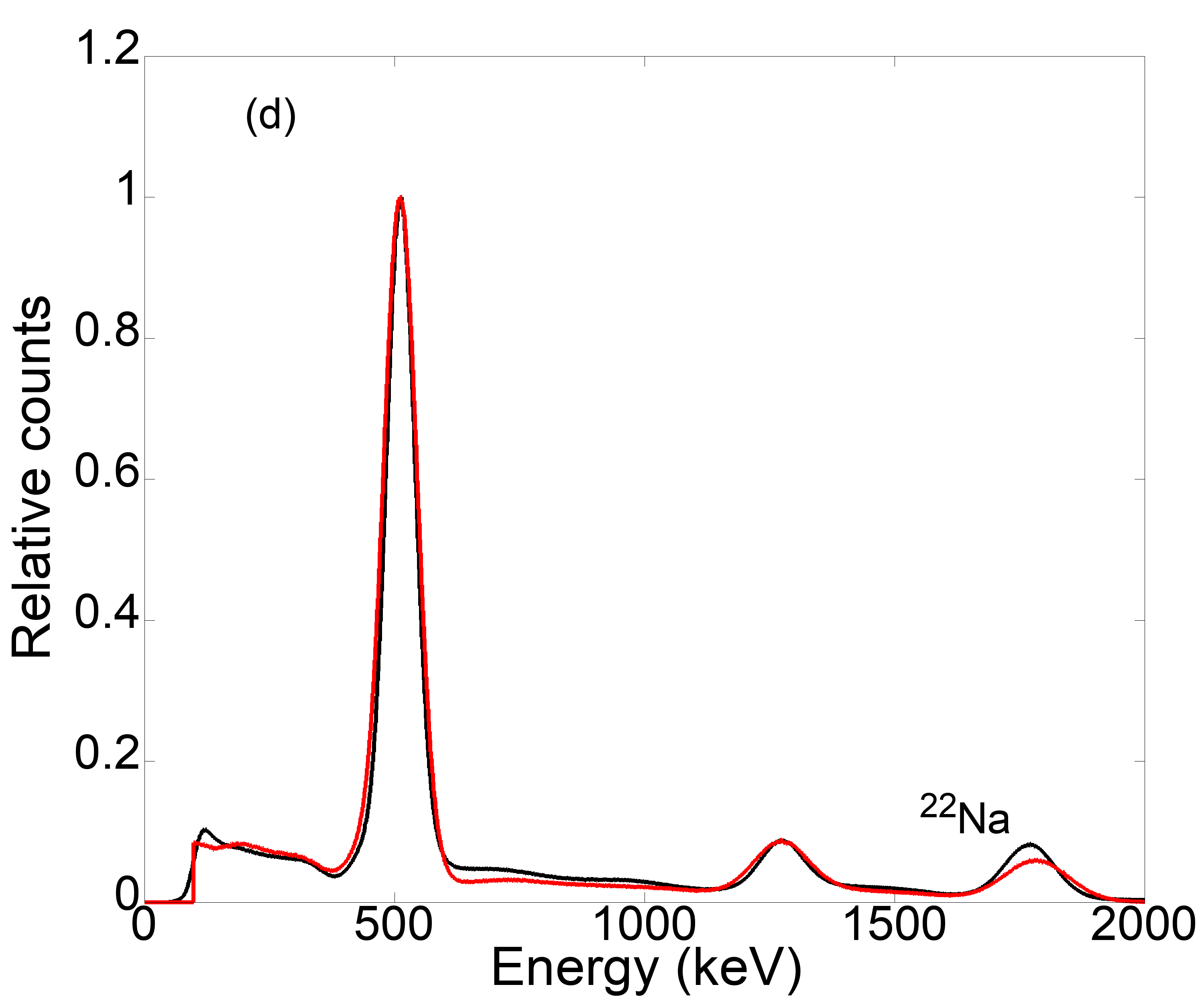}
\includegraphics[width=0.495\linewidth]{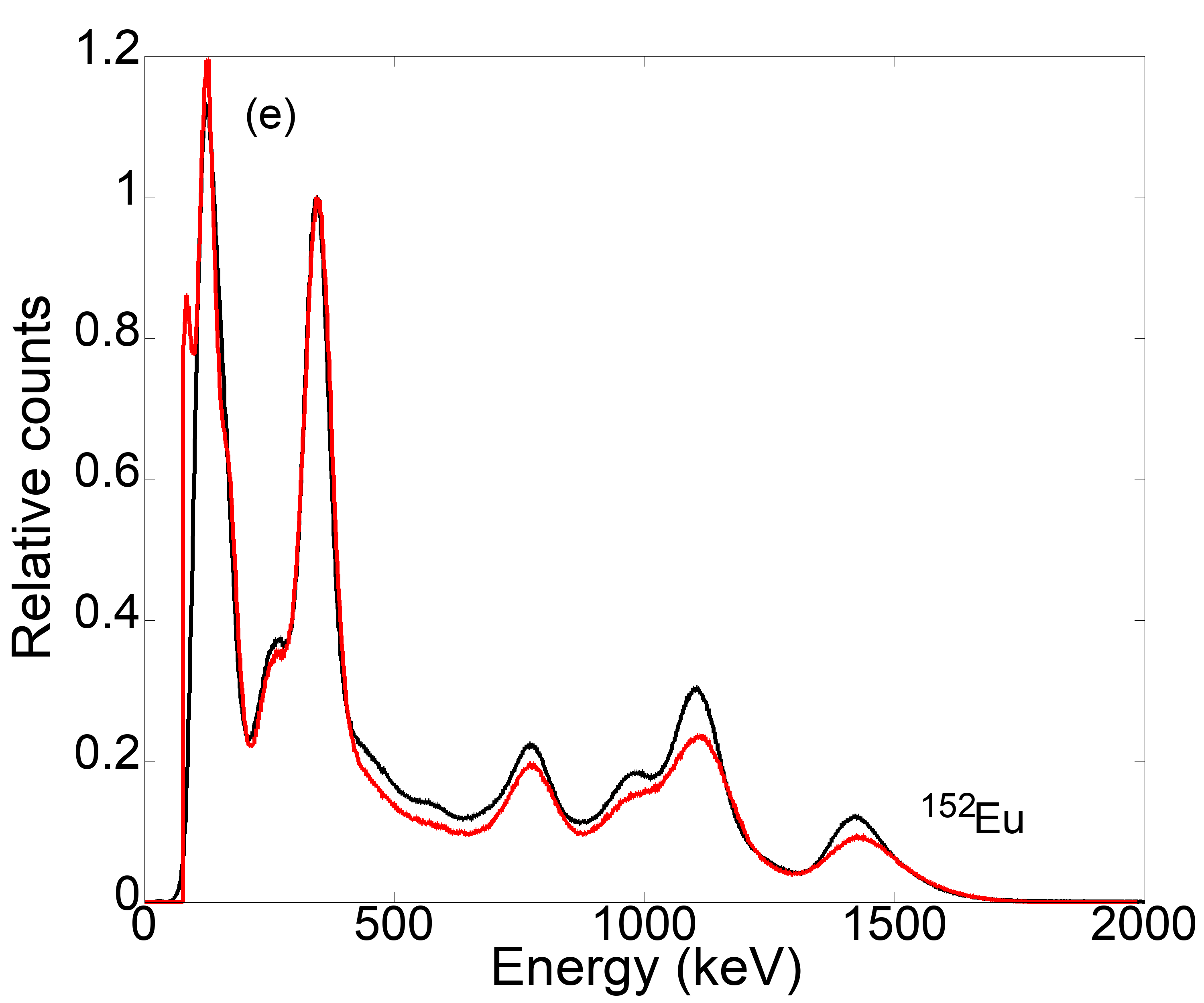}
\caption{Experimental (black line) and simulated (red line) $^{109}$Cd (a), $^{57}$Co (b), $^{137}$Cs (c), $^{22}$Na (d), and $^{152}$Eu (e) gamma ray energy spectra for the BGO scintillation detector. The color version is available online.}
\label{fig:bgo}
\end{figure}

\begin{figure}[h!]
\includegraphics[width=0.495\linewidth]{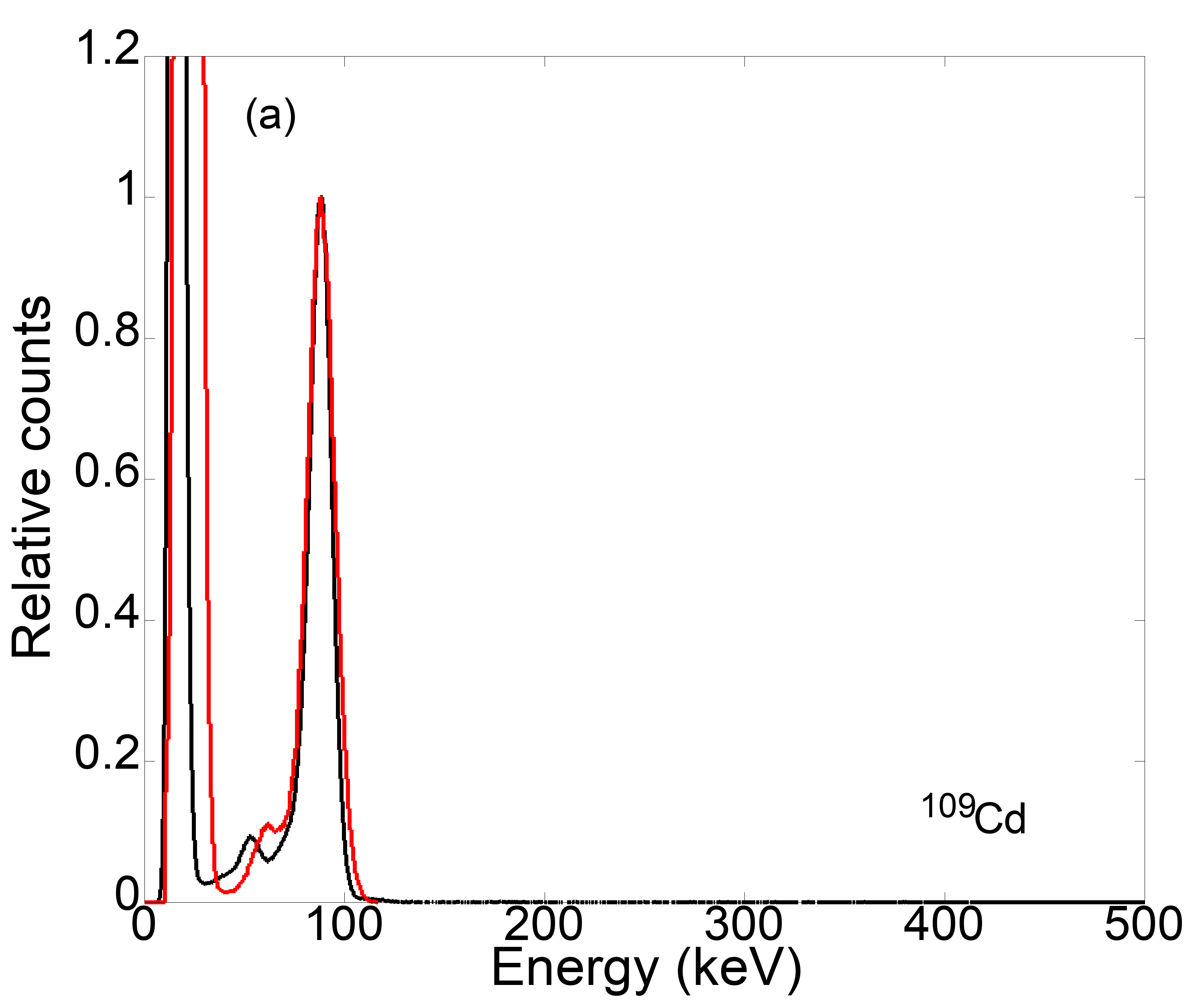}
\includegraphics[width=0.495\linewidth]{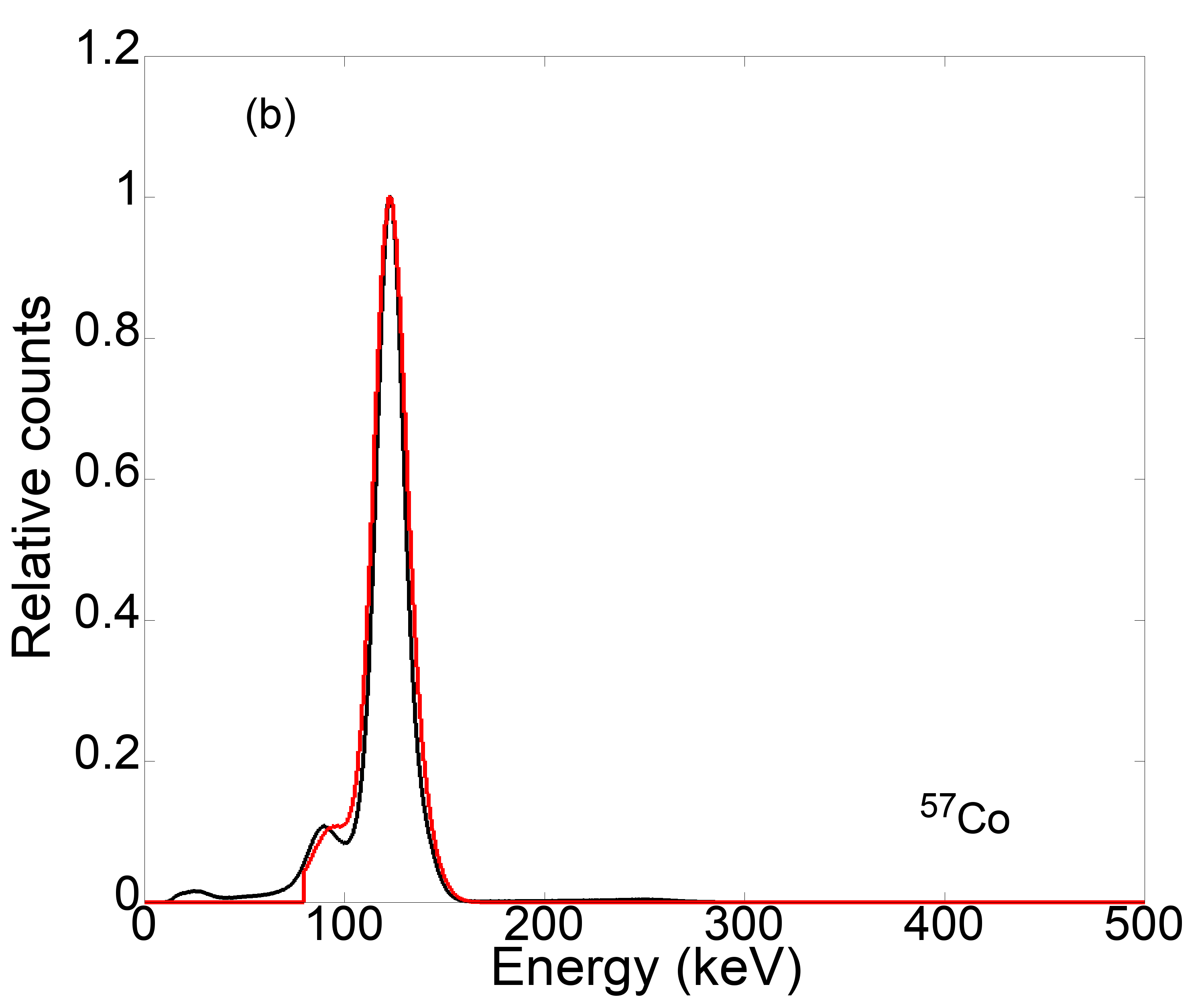}
\centerline{\includegraphics[width=0.495\linewidth]{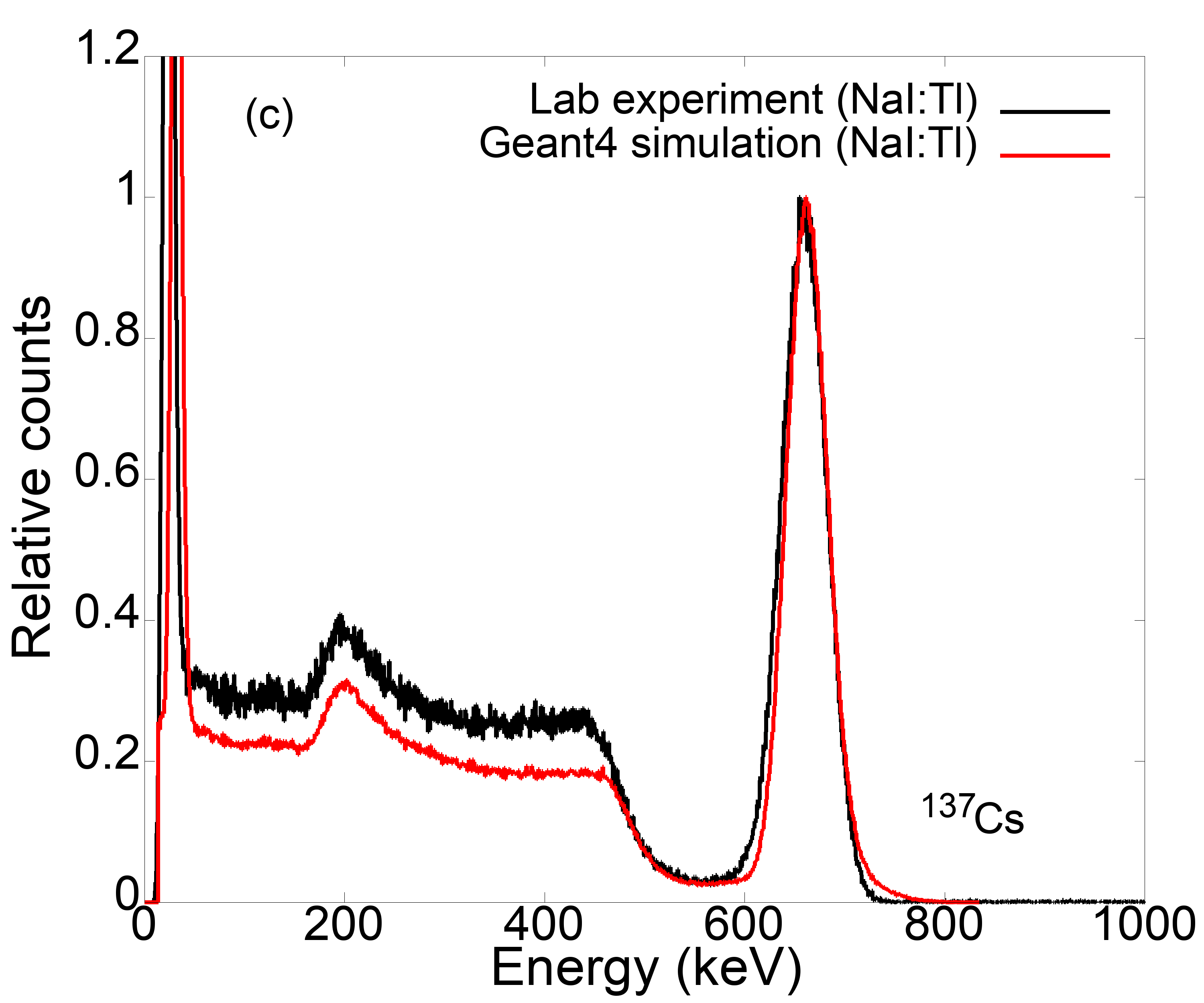}}
\includegraphics[width=0.495\linewidth]{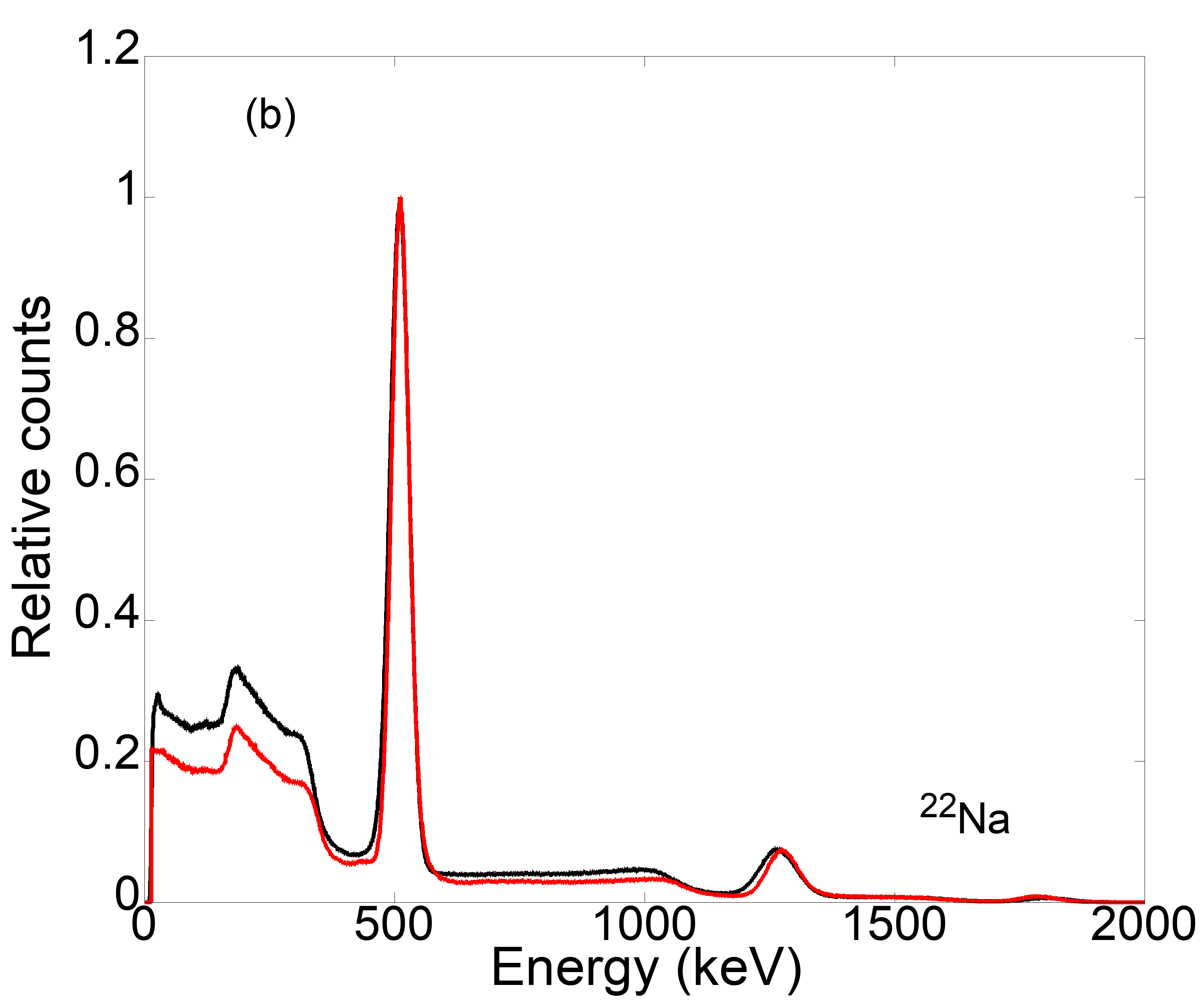}
\includegraphics[width=0.495\linewidth]{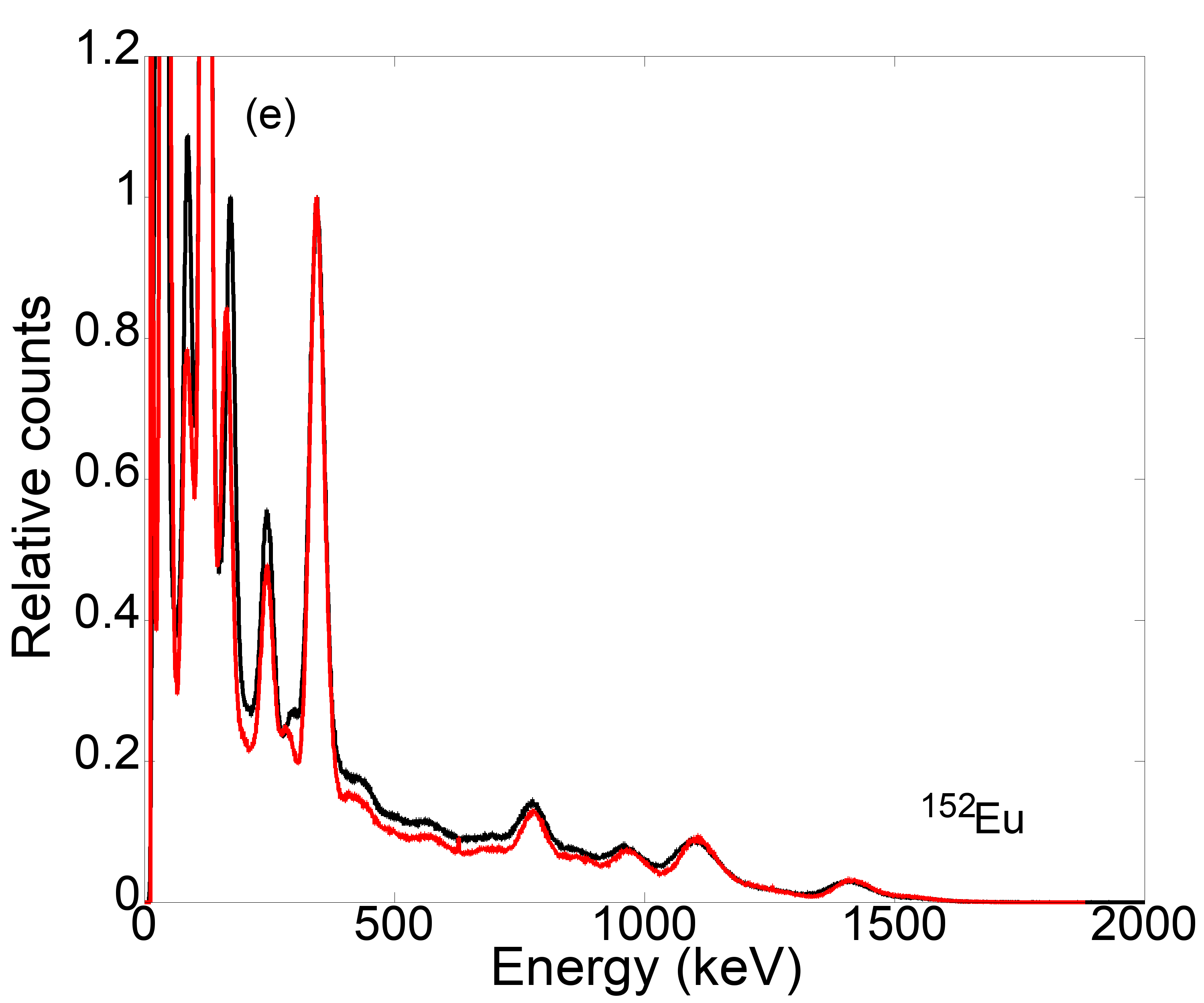}
\caption{Experimental (black line) and simulated (red line) $^{109}$Cd (a), $^{57}$Co (b), $^{137}$Cs (c), $^{22}$Na (d), and $^{152}$Eu (e) gamma ray energy spectra for the NaI:Tl scintillation detector. The color version is available online.}
\label{fig:nai}
\end{figure}

\begin{figure}[h!]
\centering
\includegraphics[width=\linewidth]{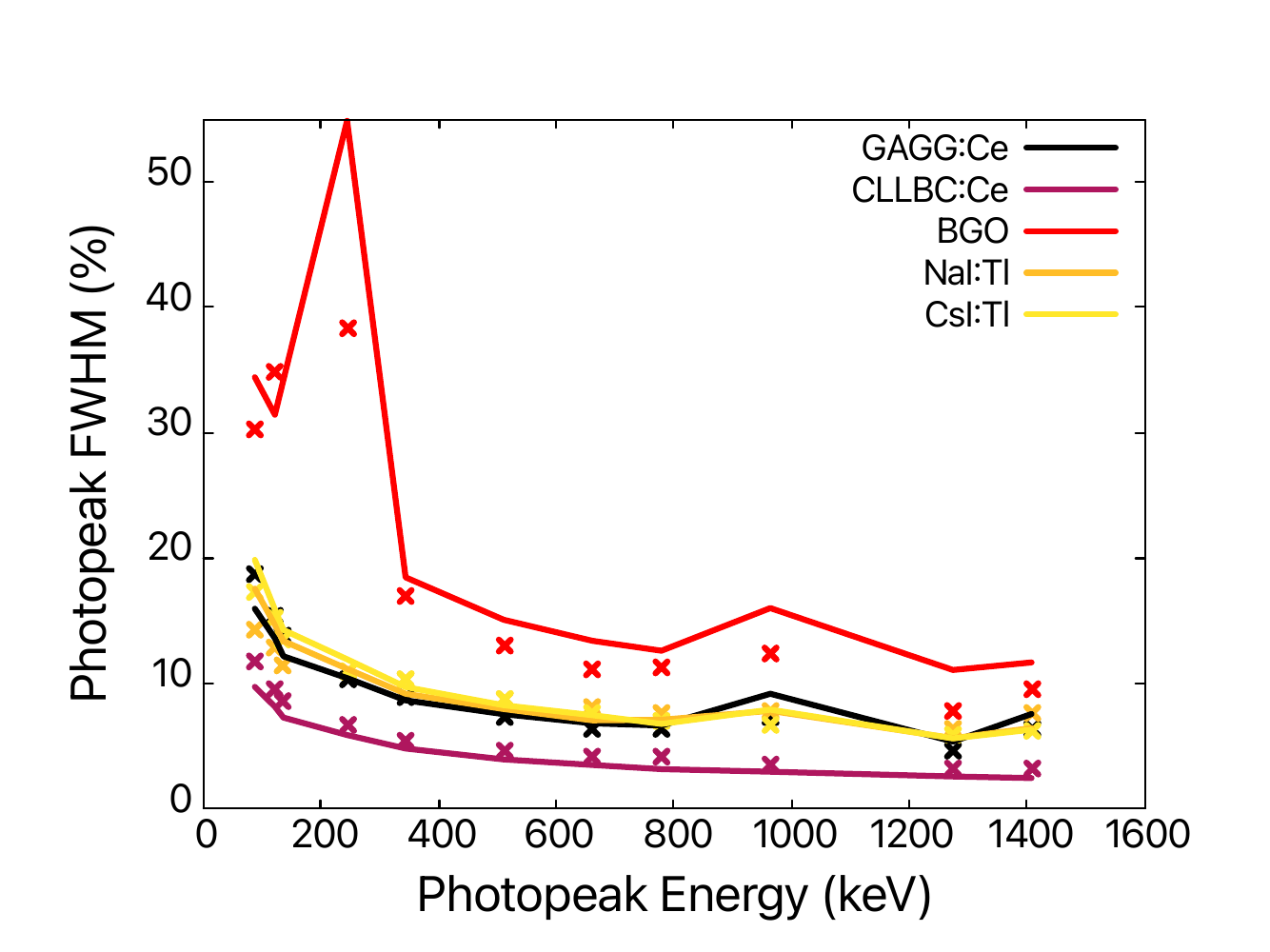}
\caption{Experimental (x-mark) and simulated (solid line) photopeak full width half maximum (FWHM) extracted from the energy spectra. The spike at 344.28 keV for BGO is caused by poor energy resolution blurring it with the other $^{152}$Eu photopeak (244.7 keV). This can be observed in Fig.\ (\ref{fig:bgo}e).}
\label{fig:spectra_fwhm}
\end{figure}

\begin{table}[h!]
\caption{NCCC between experimental and simulated energy spectra. Highlighted cells indicate acceptable NCCC values within an uncertainty of $\pm 0.004$.}\label{table:spectra_nccc}
\setlength{\tabcolsep}{6.4pt}
\begin{tabular}{|c||c|c|c|c|c|}
\hline
\makecell[c]{Radioactive \\ source} & \makecell[c]{GAGG:Ce} & \makecell[c]{CLLBC:Ce} & \makecell[c]{BGO} & \makecell[c]{NaI:Tl} & \makecell[c]{CsI:Tl} \\
\hline
$^{109}$Cd&\cellcolor{green!25}0.987&0.264&0.750&0.452&\cellcolor{green!25}0.998\\
\hline
$^{57}$Co&\cellcolor{green!25}0.996&\cellcolor{green!25}0.993&\cellcolor{green!25}0.998&\cellcolor{green!25}0.994&\cellcolor{green!25}0.993\\
\hline
$^{137}$Cs&\cellcolor{green!25}0.987&0.977&\cellcolor{green!25}0.987&0.824&\cellcolor{green!25}0.997\\
\hline
$^{22}$Na&\cellcolor{green!25}0.996&\cellcolor{green!25}0.991&\cellcolor{green!25}0.995&\cellcolor{green!25}0.990&\cellcolor{green!25}0.996\\
\hline
$^{152}$Eu&0.965&0.959&0.972&0.823&\cellcolor{green!25}0.987\\
\hline
\end{tabular}
\end{table}

\section{Results and Discussion}
\begin{figure}[h!]
\includegraphics[width=0.495\linewidth]{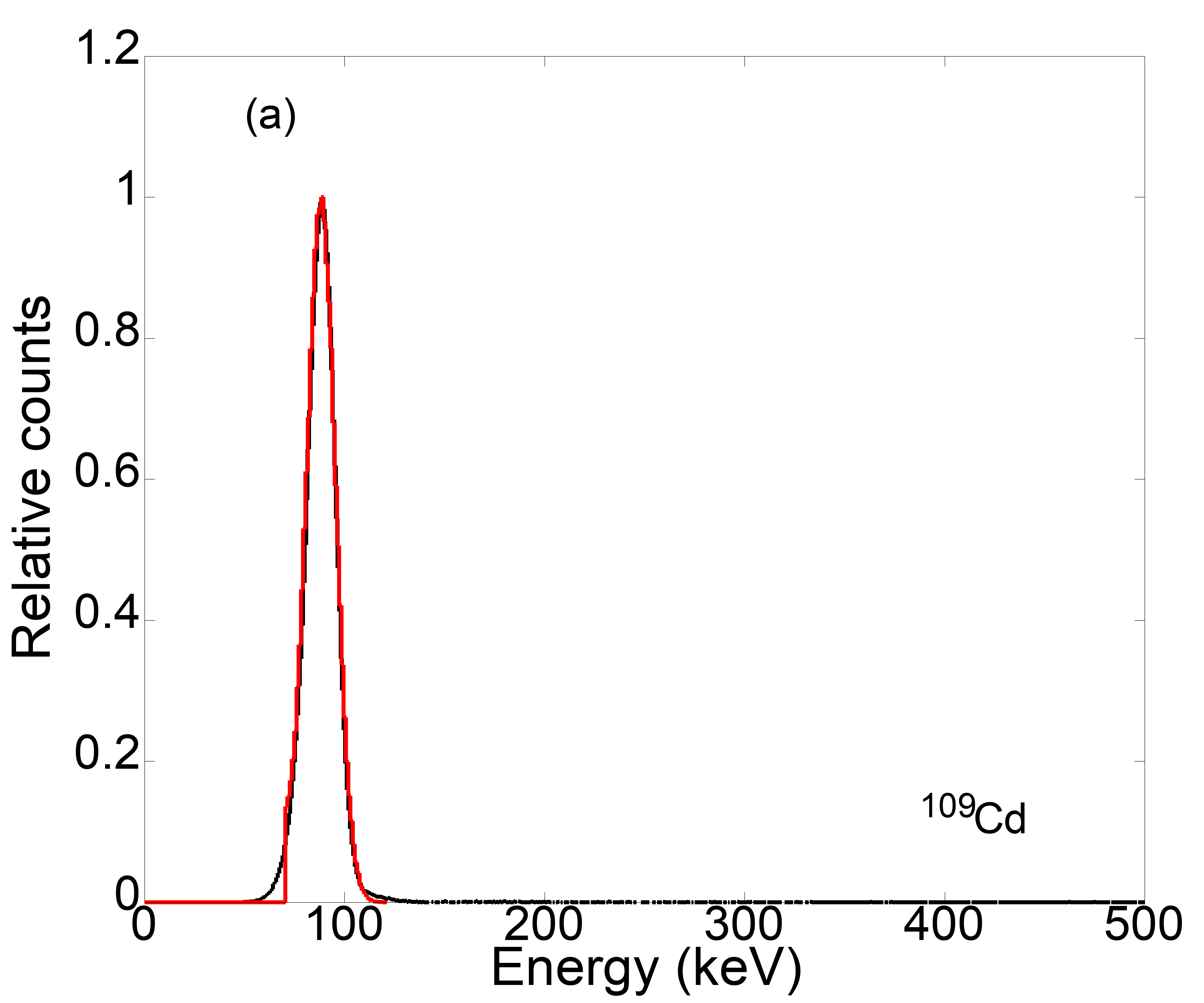}
\includegraphics[width=0.495\linewidth]{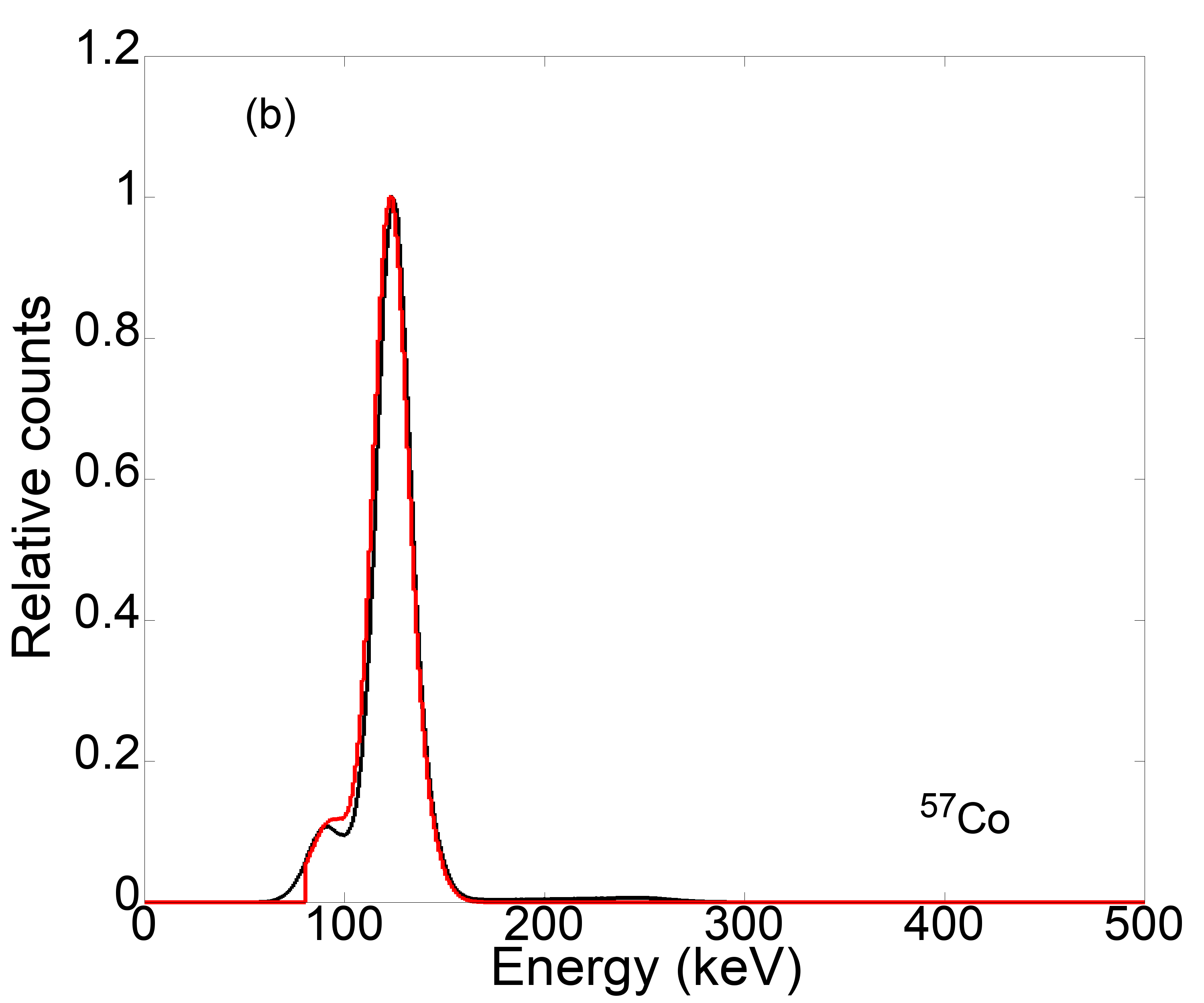}
\centerline{\includegraphics[width=0.495\linewidth]{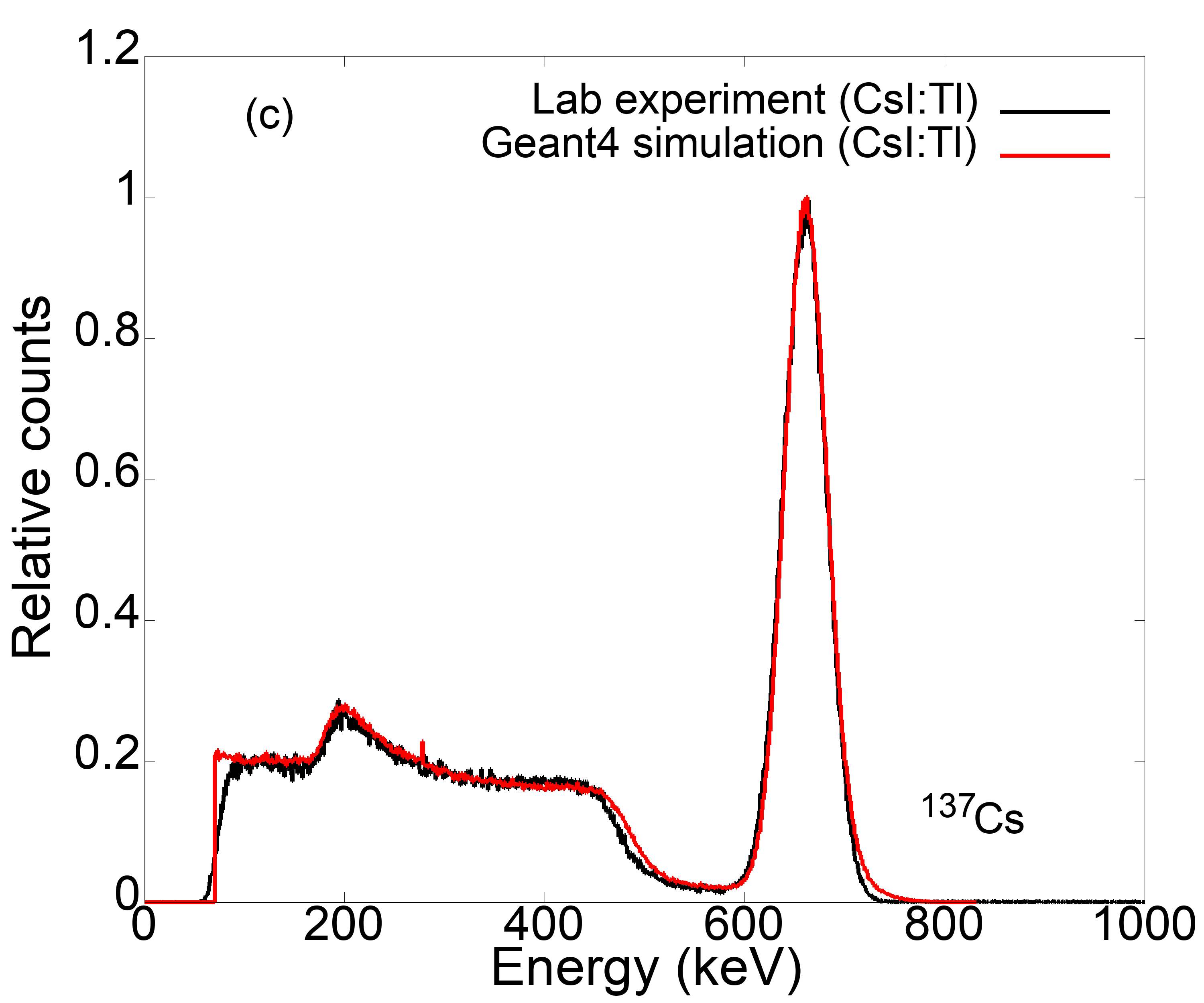}}
\includegraphics[width=0.495\linewidth]{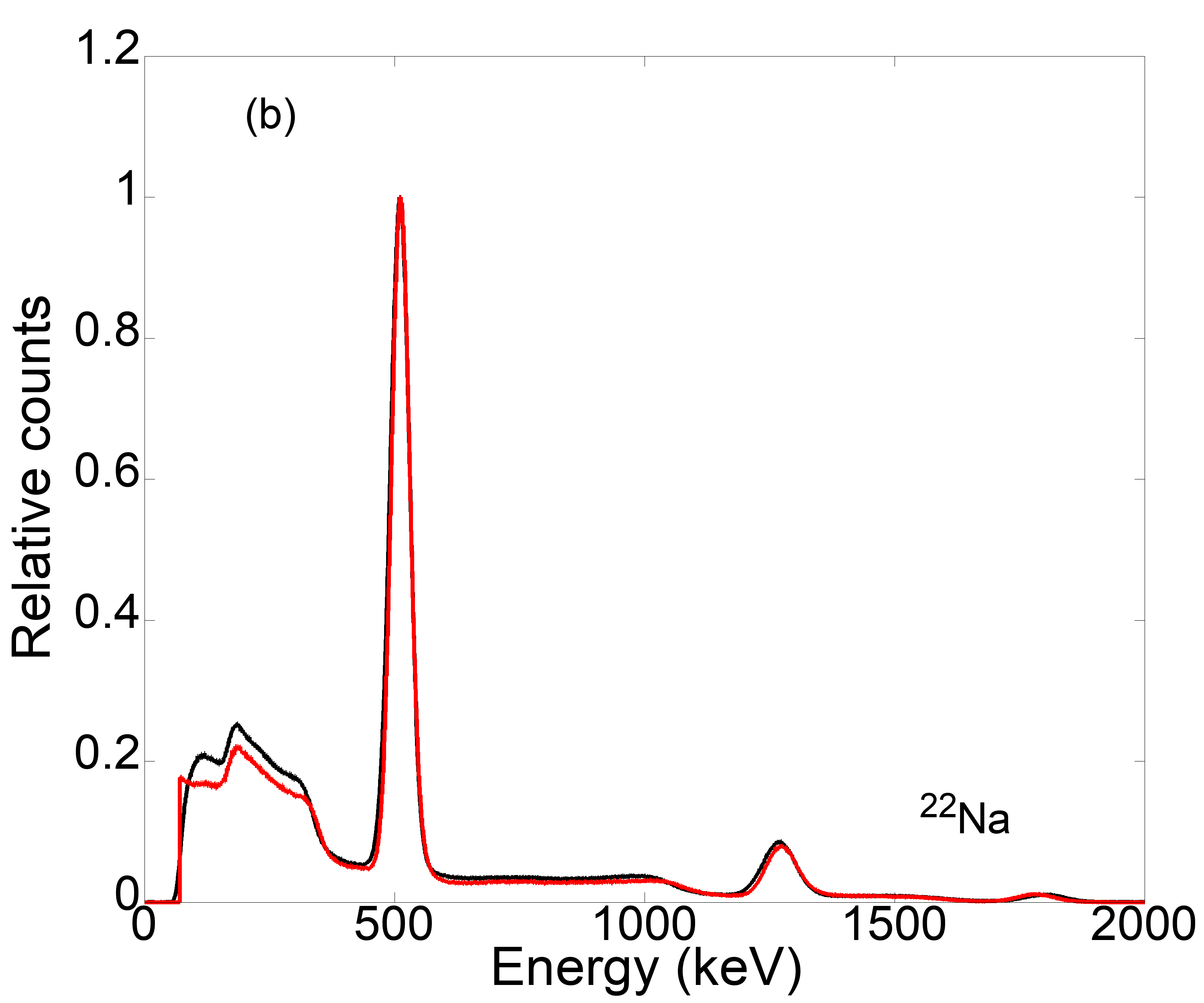}
\includegraphics[width=0.495\linewidth]{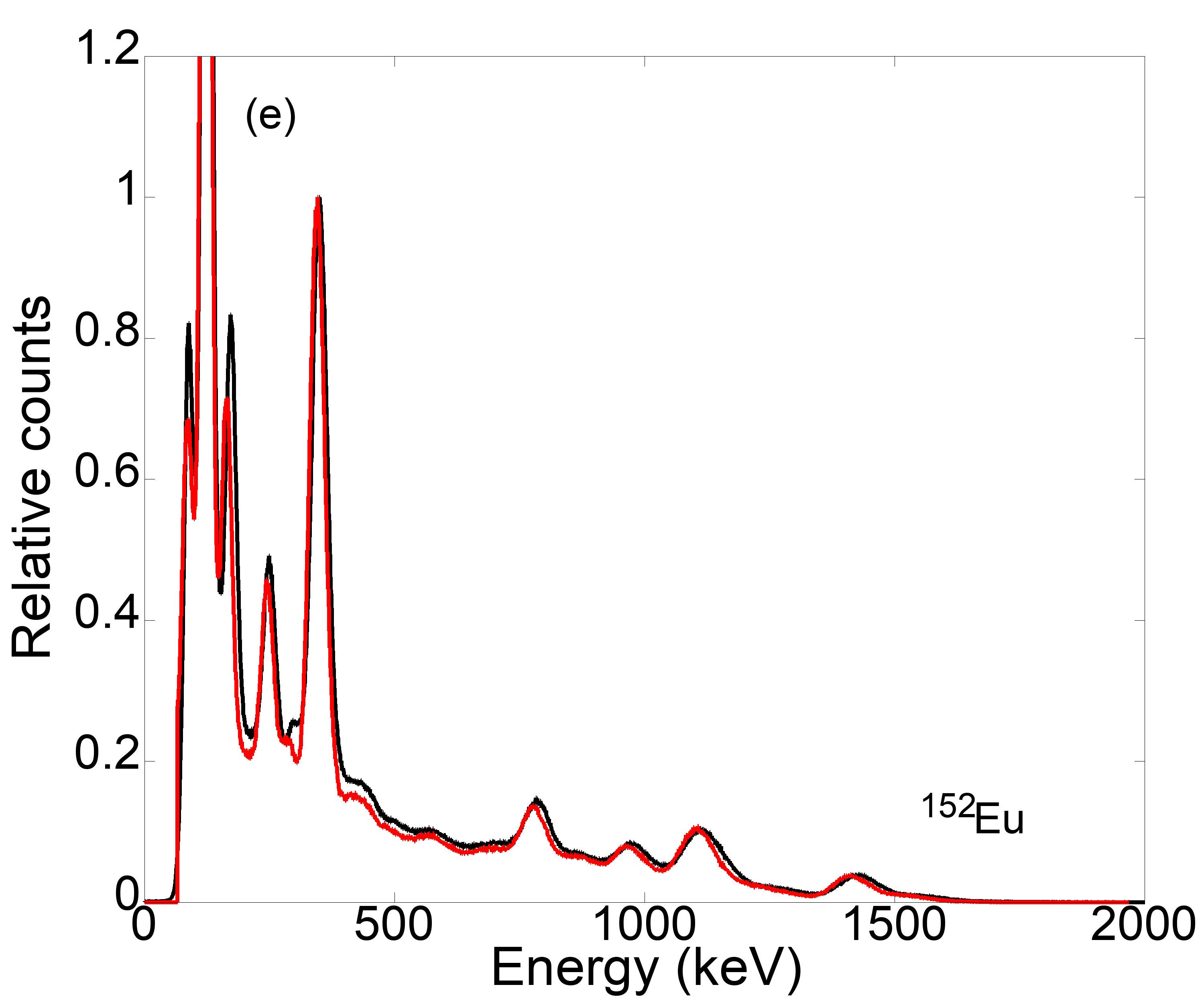}
\caption{Experimental (black line) and simulated (red line) $^{109}$Cd (a), $^{57}$Co (b), $^{137}$Cs (c), $^{22}$Na (d), and $^{152}$Eu (e) gamma ray energy spectra for the CsI:Tl scintillation detector. The color version is available online.}
\label{fig:csi}
\end{figure}

Figs.\ (\ref{fig:gagg}) through (\ref{fig:csi}) present the simulated and experimental energy spectra for the five scintillation materials. Overall, the Geant4 application accurately predicted the spectral features observed in the experimental energy spectra. These features included the Compton continuum,  edge, photopeaks, annihilation peaks, double-sum peaks, and, in cases with sufficient low-level detection, X-ray peaks. The difference between simulated and experimental photopeak FWHM for GAGG:Ce, CLLBC:Ce, NaI:Tl, and CsI:Tl above 88 keV was less than 2\%, as shown in Fig.\ (\ref{fig:spectra_fwhm}). Moreover, the majority of NCCC values in Table (\ref{table:spectra_nccc}) demonstrated agreement between the experimental and simulated detector platforms within an uncertainty of $\pm0.004$. This uncertainty was calculated by propagating the uncertainties from the experimental and simulated spectra counts using the partial derivatives of the NCCC function in (\ref{eq:nccc}). In this section, the performance of the simulation platform is discussed in more detail by comparing spectral features and figures of merit to the energy spectra generated from the experimental platform. 

The experimental $^{22}$Na spectra for GAGG:Ce and BGO revealed a double-sum peak at 1766 keV and 1769 keV (see Figs.\ (\ref{fig:gagg}d) and (\ref{fig:bgo}d)). This was caused by the 511 keV and 1275 keV gamma ray interacting with the SiPM in the same time window. As the double-sum peaks were not 1786 keV (511 keV + 1275 keV), the SiPM did not fully capture the optical response from the scintillator crystals. Double-sum peaks were negligible in the $^{22}$Na energy spectra for CLLBC:Ce, NaI:Tl, and CsI:Tl (Figs.\ (\ref{fig:cllbc}d), (\ref{fig:nai}d), and (\ref{fig:csi}d)) due to their smaller attenuation coefficients, making it less likely for both gamma rays to interact with them in the same time window. GAGG:Ce and BGO's larger attenuation coefficients (see Table (\ref{table:scintillator_crystal_types})) corresponded to less statistical noise in the $^{137}$Cs spectra (Figs.\ (\ref{fig:gagg}c) and (\ref{fig:bgo}c)) compared to the other scintillators. Statistical noise was more prevalent for $^{137}$Cs compared to other isotopes because $^{137}$Cs had the lowest radioactivity. The peak immediately above the low-level detection threshold observed in the $^{137}$Cs and $^{22}$Na experimental spectra (e.g., GAGG:Ce in Figs.\ (\ref{fig:gagg}c) and (\ref{fig:gagg}d)) was attributed to electronic noise. This peak was not observed in the simulated spectra because electronic noise was not accounted for in the simulation platform.

The average difference between simulated and experimental photopeak FWHM was $<$1\% for CsI:Tl, 1\% for GAGG:Ce and CLLBC:Ce, 2\% for NaI:Tl, and 4\% for BGO. For each scintillator, the energy resolution generally improved as the characteristic gamma ray energy increased. Although this was true for $^{152}$Eu within the photopeaks for the same source, it was not true between photopeaks of other isotopes. For example, the 344 keV and 964 keV $^{152}$Eu photopeaks for CsI:Tl had a FWHM of 9.72\% and 7.92\% respectively, however, the FWHM for the 662 keV photopeak was 7.47\%. This phenomenon was attributed to cascade summing from multiple $^{152}$Eu radioactive decays. The poor energy resolution of BGO blurred the 122 keV and 136 keV $^{57}$Co photopeaks together so that the 136 keV photopeak could not be resolved, as shown in Fig.\ (\ref{fig:bgo}b). Moreover, the poor energy resolution of BGO was attributed to the 16\% difference between the simulated and experimental 245 keV $^{152}$Eu photopeak FWHM. Overall, CLLBC:Ce had the best energy resolution (lowest photopeak FWHM) across the 30 keV to 2 MeV energy range (see Fig.\ (\ref{fig:spectra_fwhm})). For the simulated 662 keV photopeak, CLLBC:Ce had an energy resolution of 3.51\%, GAGG:Ce with 6.83\%, NaI:Tl with 7.07\%, CsI:Tl with 7.47\%, and BGO with 13.4\%.

Table (\ref{table:spectra_nccc}) highlights the acceptable NCCC values within the uncertainty of $\pm$0.004. The NCCC values below 0.99 were attributed to discrepancies between the peak-to-Compton ratio of the simulated and experimental energy spectra and, for BGO, scintillator non-linearity. The non-linear effect was observed in the $^{109}$Cd and $^{152}$Eu experimental energy spectra with BGO at low energies (see Fig.\ (\ref{fig:bgo}a) and (\ref{fig:bgo}e)), where the corresponding NCCC values were 0.750 and 0.972. This was not unexpected as non-linearity was not accounted for in the BGO energy calibration process as the 88 keV $^{109}$Cd photopeak coincided with the low-level detection threshold in the experiments. Discrepancies in the peak-to-Compton ratios were observed in multiple energy spectra, such as the $^{152}$Eu energy spectrum with CLLBC:Ce (see Fig.\ (\ref{fig:cllbc}e)) and $^{137}$Cs with NaI:Tl (see Fig.\ (\ref{fig:nai}c)). They were also observed in energy spectra with acceptable NCCC values, such as the $^{22}$Na spectra with NaI:Tl and CsI:Tl (see Figs.\ (\ref{fig:nai}d) and (\ref{fig:csi}d)). These discrepancies were caused by different amounts of Compton scattering in the simulations and experiments, which was demonstrated in particular by the NaI:Tl and CsI:Tl simulations as they only differed by the presence of sodium or caesium in the scintillators. Since $Z_{\text{eff}}$ was lower for NaI:Tl compared to CsI:Tl (see Table (\ref{table:scintillator_crystal_types})), the gamma ray was more likely to Compton scatter out of the NaI:Tl and interact with the detector and surrounding environment. Different amounts of Compton scattering between the simulations and experiments were attributed to lower fidelity electronic modelling, geometry approximations made between the two internal PCBs in Fig.\ (\ref{fig:complete_housing}b), and material composition approximations of the table and surrounding lab environment (see Table (\ref{table:geant4_geometries})).

\section{Conclusion}
A Geant4 optical material parameter set for GAGG:Ce, CLLBC:Ce, BGO, NaI:Tl, and CsI:Tl was compiled and used to help validate a detailed SiPM-based scintillation detector simulation platform developed in Geant4. The response of these scintillator crystals to characteristic gamma rays with energies between 30 keV to 2 MeV were simulated and compared to experimental measurements. The simulation platform successfully predicted the spectral features measured in the experiments. Moreover, all photopeaks above 88 keV simulated for GAGG:Ce, CLLBC:Ce, NaI:Tl, and CsI:Tl had a FWHM within 2\% of the experimental value, and the majority of NCCC values indicated agreement between the simulated and experimental energy spectra. There were minor discrepancies in the simulated FWHM and NCCC values, which were attributed to: 1) variations in the Compton continuum caused by approximations within the detector housing and surrounding lab environment, and 2) detector signal processing electronics modelling. This scintillator parameter set will enable efficient development of scintillator technology in space, medical imaging, homeland security, and environmental monitoring applications.

\begin{figure*}[h!]
\includegraphics[width=0.495\linewidth]{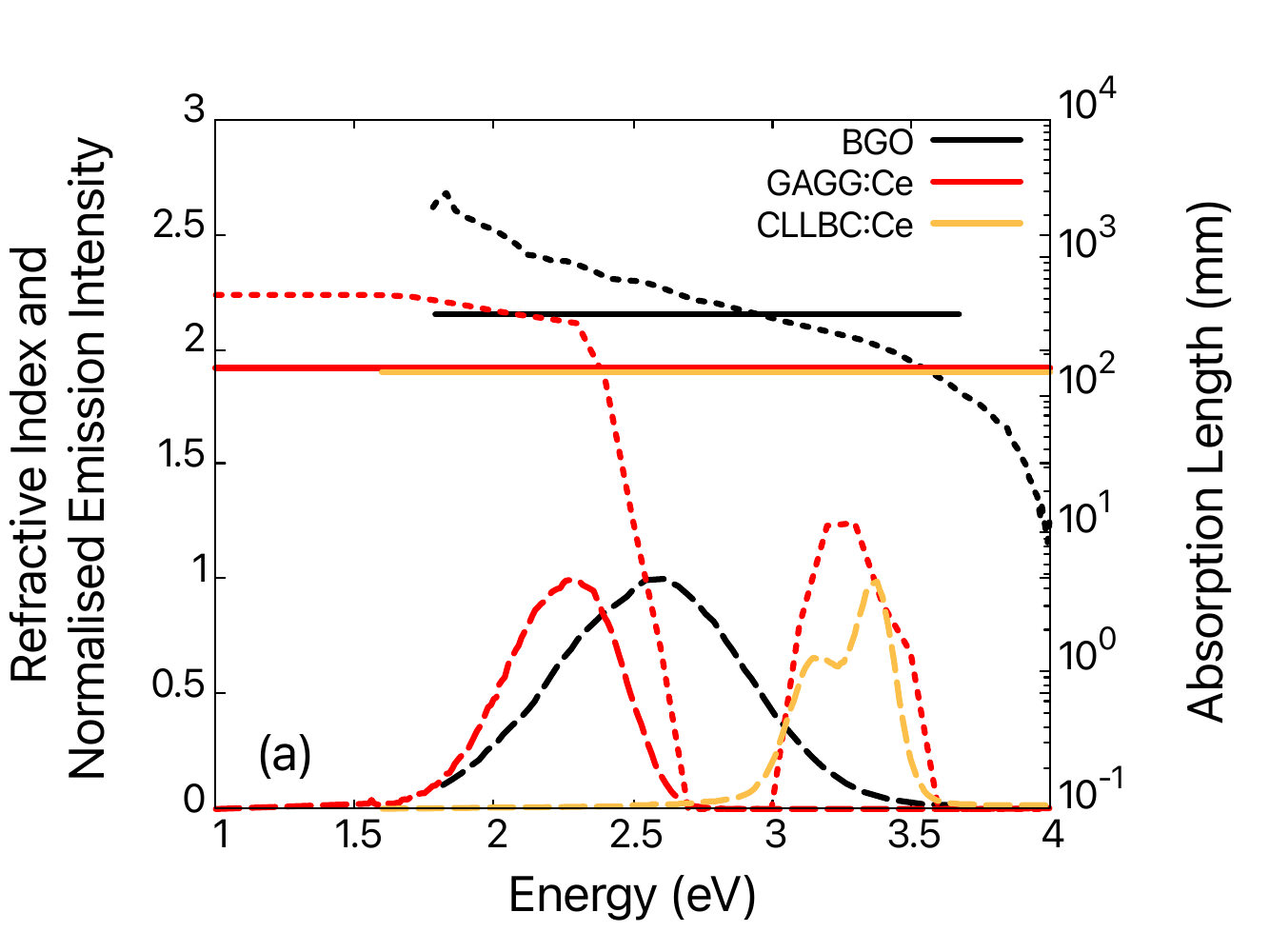}
\includegraphics[width=0.495\linewidth]{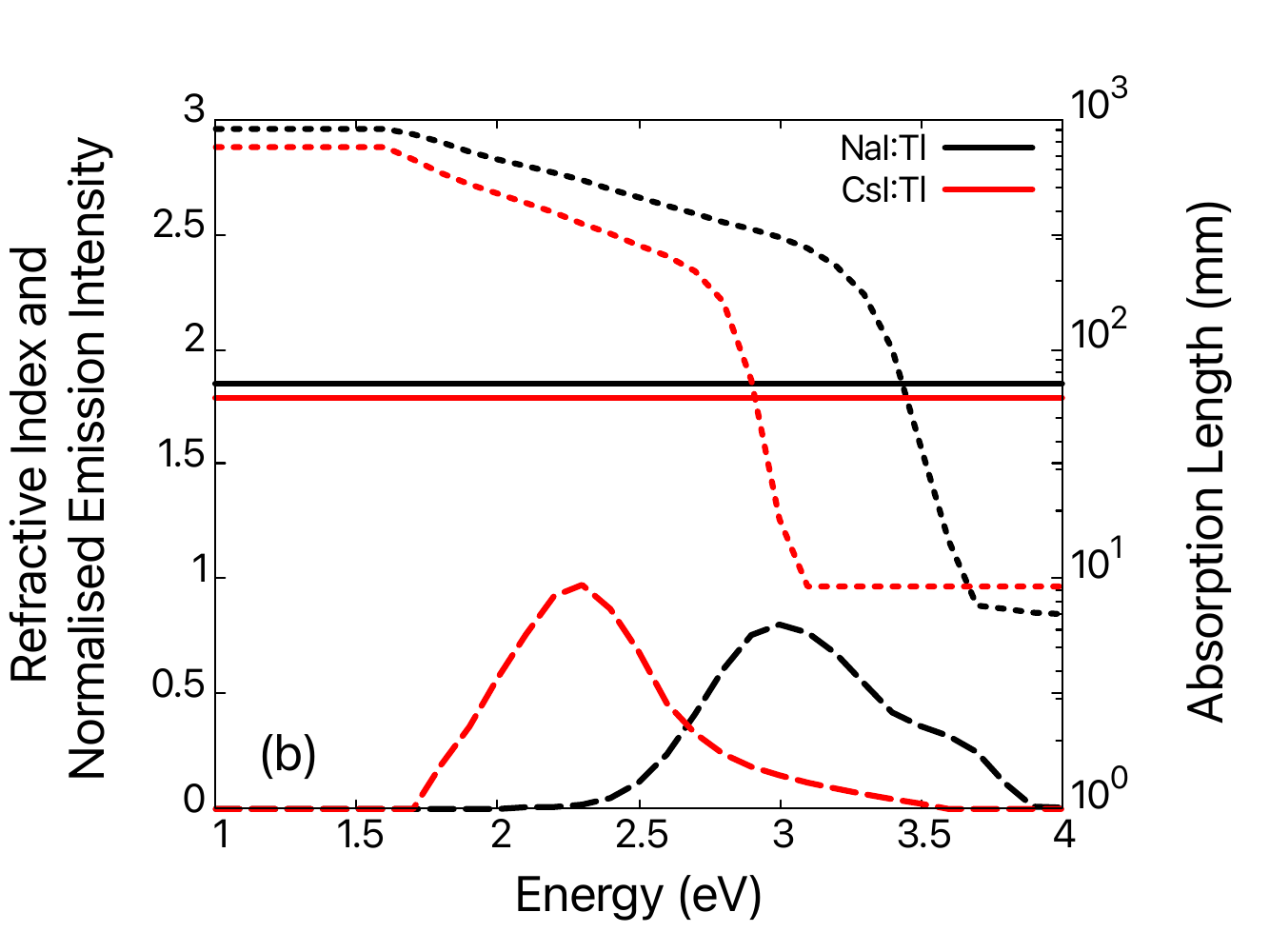}
\includegraphics[width=0.495\linewidth]{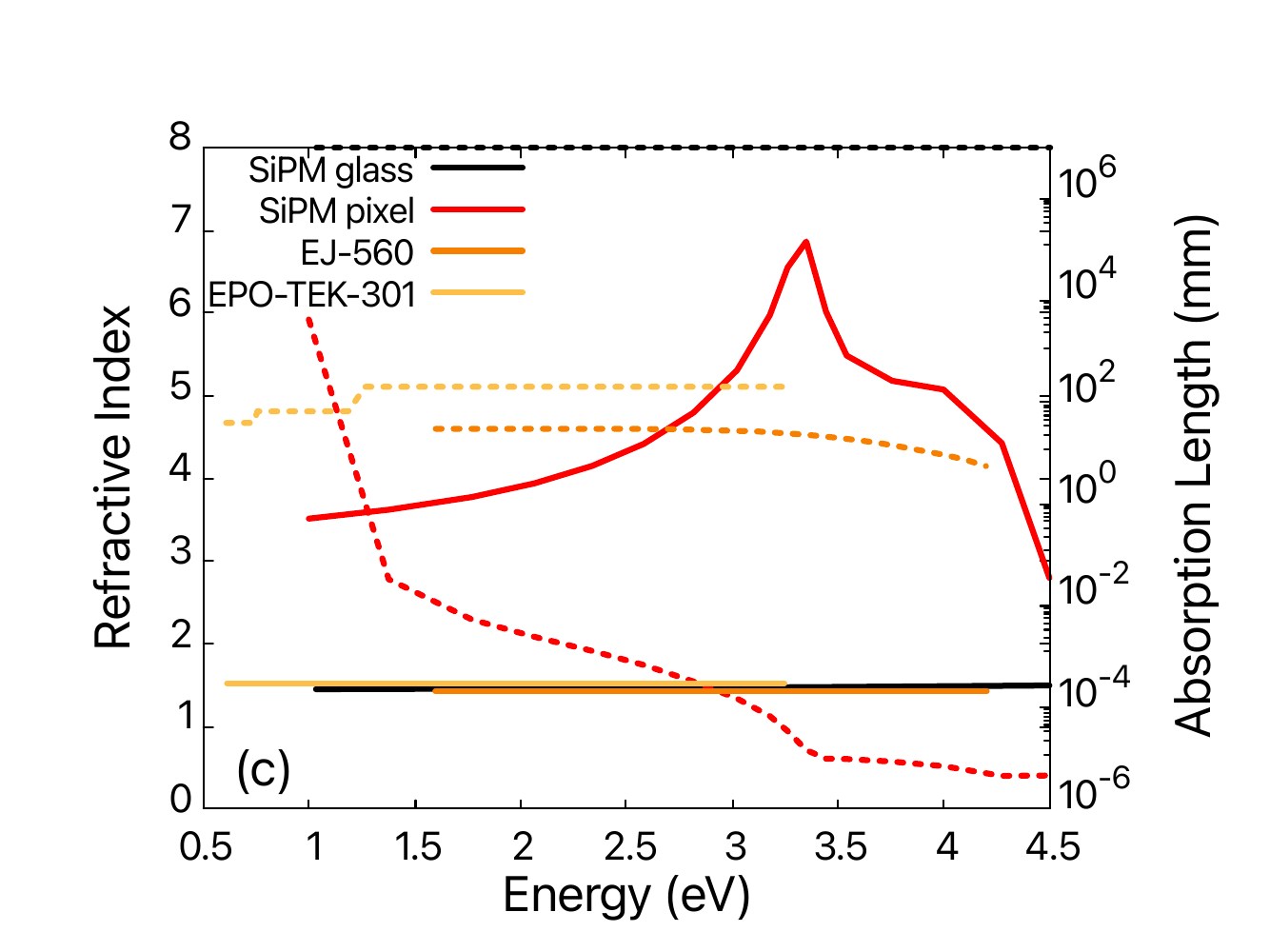}
\includegraphics[width=0.495\linewidth]{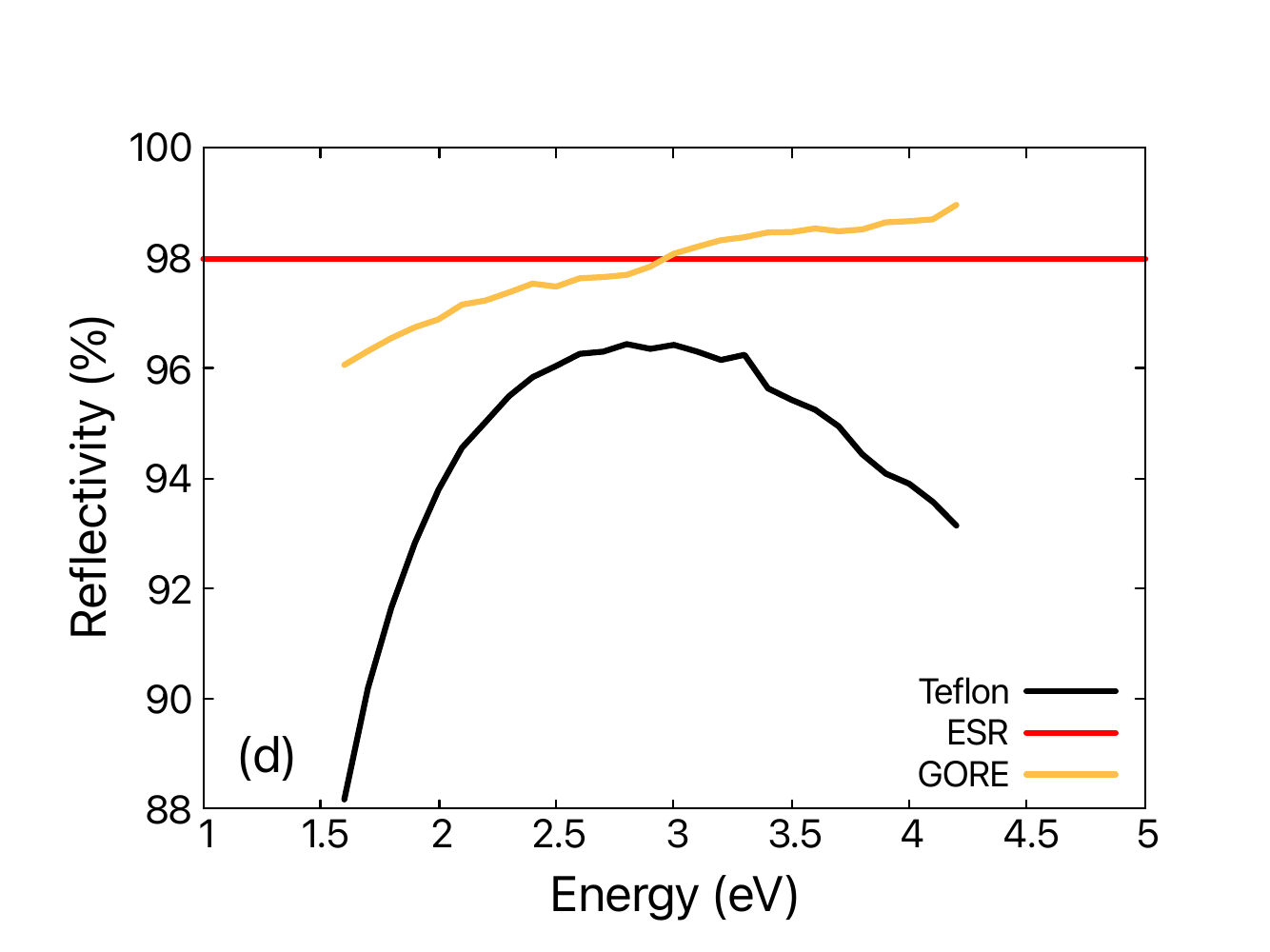}
\caption{Material refractive index and reflectivity (solid line), normalised emission intensity (dashed line), and absorption length (dotted line) data implemented in the Geant4 simulation platform. (a) Refractive index, emission intensity, and absorption length for GAGG:Ce \citep{Brown2023a,Brown2021,Kobayashi2012} and BGO \citep{Mao2008, EPICBGO2023}. Refractive index and emission intensity for CLLBC:Ce \citep{Brown2023,Shirwadkar2012}. (b) Refractive index, emission intensity, and absorption length for NaI:Tl and CsI:Tl \citep{Brown2023a,Brown2021,Mao2008}. (c) Refractive index and absorption length for the glass \citep{Brown2020}, SiPM pixel \citep{Philipp1960}, EJ-560 \citep{EJ5602021}, and EPO-TEK-301 \citep{EPOTEK3012013}. (d) Reflectivity for the Teflon tape \citep{Janecek2012}, ESR \citep{3M2020}, and GORE diffuse reflector \citep{Janecek2012}.}
\label{fig:material_property_data}
\end{figure*}

\appendix

\section*{Geant4 Material Properties}
This appendix contains the refractive index and reflectivity, emission spectrum, and absorption length reference data for all optical materials used in the Geant4 simulation platform. The data for GAGG:Ce, CLLBC:Ce, BGO, NaI:Tl, and CsI:Tl is presented in Figs.\ (\ref{fig:material_property_data}a) and (\ref{fig:material_property_data}b). The EPO-TEK-301, Teflon tape, ESR, GORE diffuse reflector, glass, SiPM pixel, and EJ-560 optical pad reference data is displayed in Figs.\ (\ref{fig:material_property_data}c) and (\ref{fig:material_property_data}d). All optical datasets from Figs.\ (\ref{fig:material_property_data}a) through (\ref{fig:material_property_data}d) are available through the IEEE Dataport.

\section*{Acknowledgment}
L.\ Miller would like to acknowledge the Optical Sciences Centre at Swinburne University of Technology for support of the project.

\bibliographystyle{IEEEtran}
\bibliography{export}

\end{document}